\documentclass[aps,10pt,superscriptaddress,showpacs]{revtex4}

\usepackage{amsfonts}
\usepackage{amsmath}
\usepackage{amssymb}
\usepackage{graphicx}
\usepackage{epsfig}

\begin{document}

\title{Interaction of fast charged projectiles with two-dimensional electron
gas: Interaction and disorder effects}
\author{Hrachya B. Nersisyan}
\email{hrachya@irphe.am}
\affiliation{Division of Theoretical Physics, Institute of Radiophysics and Electronics,
Alikhanian Brothers Street 1, 378410 Ashtarak, Armenia}
\affiliation{Centre of Strong Fields Physics, Yerevan State University, Alex Manoogian
str. 1, 375025 Yerevan, Armenia}
\author{Amal K. Das}
\email{akdas@dal.ca}
\affiliation{Department of Physics, Dalhousie University, Halifax, Nova Scotia B3H 3J5,
Canada}
\date{\today}

\begin{abstract}
The results of a theoretical investigation on the stopping power of ions
moving in a disordered two-dimensional degenerate electron gas are
presented. The stopping power for an ion is calculated employing linear
response theory using the dielectric function approach. The disorder, which
leads to a damping of plasmons and quasiparticles in the electron gas, is
taken into account through a relaxation time approximation in the linear
response function. The stopping power for an ion is calculated in both the
low- and high-velocity limits. In order to highlight the effects of damping
we present a comparison of our analytical and numerical results, in the case
of point-like ions, obtained for a non-zero damping with those for a
vanishing damping. It is shown that the equipartition sum rule first
formulated by Lindhard and Winther for three-dimensional degenerate electron
gas does not necessarily hold in two-dimensions. We have generalized this
rule introducing an effective dielectric function. In addition some new
results for two-dimensional interacting electron gas have been obtained. In
this case the exchange-correlation interactions of electrons are considered
via local-field-corrected dielectric function.
\end{abstract}

\maketitle

\section{Introduction}
\label{sec:intr}

There is an ongoing interest in the theory of interaction of swift charged
projectiles with condensed matter. Although most theoretical works have
reported on the energy loss of ions in a target medium which is modelled as
a three-dimensional (3D) electron gas, the two-dimensional (2D) case has not
yet received as much attention as the 3D case. A 2D electron system is now
experimentally realizable in a laboratory. In the last three decades or so
many interesting and intriguing properties of a 2D electron gas have been
explored. For a recent update on some of these developments we refer to
Refs.~\cite{dur04,fun07}. A widely used 2D electron system is realized at
the interface between GaAs and Ga$_{1-x}$Al$_{x}$As, and in the interface
metal-oxide-semiconductors (MOS). The interaction of charged particles with
an electron gas is an important probe of many-body interactions in the
target electron medium. It is known that many-body properties of an electron
gas vary in notable aspects with spatial dimensions. It is therefore of
interest to make a detailed study of interaction of charged particles with a
2D electron gas. This theoretical study is also of relevance to device
applications e.g. in using ion implantation in devices which involve 2D
electron systems.

In general, interaction of charged projectiles with condensed matter can be
studied by means of the stopping power (SP) of the target medium. The SP
accounts for the energy loss by an external charged projectile as it passes
through and interacts with matter. And the SP of a medium can be used to
construct diagnostic tools for studying this kind of physical systems. There
have been several theoretical approaches to the energy loss and SP for 3D
systems, and some of these approaches have been applied also to a 2D
electron gas. Among previous theoretical works on a 2D electron gas some are
based on the linear response dielectric function method \cite%
{bre93,bre94,bre98,wan95,wang95,bal06} and quantum scattering theory \cite%
{nag95,kra95,wan97,ber97}. Further works have dealt with some nonlinear
screening effects through a quadratic response approach within the
random-phase approximation \cite{ber99}, the employment of density
functional theory \cite{zar05} and in a method based on frequency moments of
the energy loss function \cite{bal06}.

In this paper we shall consider fast charged projectiles and hence a linear
response theory to calculate energy loss is expected to be adequate.
Previously, within this approach, Bret and Deutsch calculated the SP of an
ion \cite{bre93,bre94} and a dicluster \cite{bre98} in a 2D electron gas for
any degeneracy. Their results show some interesting differences with the
corresponding results for a 3D case. Of special interest is their finding
that the leading term of the asymptotic expansion of the SP in a
high-velocity limit decreases as $1/v$, where $v$ is the projectile
velocity, which differs from the well-known form predicted by the
Bethe-Bloch formula \cite{bet30,lin54,lin64} in the 3D case. The
calculations in Refs.~\cite{bre93,bre94,bre98} are based on the random-phase
approximation (RPA) which works well if electron-electron interaction can be
neglected. Now, in 2D systems, electron density can be varied. For moderate
values of electron density e.g. in semiconductors electron-electron
interaction may not be negligible and going beyond RPA is desirable.

Our objective is to consider two physically motivated aspects of a 2D
electron gas in the context of energy loss. For the first part of our study
we consider a disordered electron gas which contains impurities. The effect
of these impurities is included through a phenomenological relaxation time
for electrons due to scattering by impurities. For this disordered system we
use a linear response dielectric function in RPA and in a number-conserving
relaxation time approximation (RTA), which was first considered by Mermin
\cite{mer70} and then by Das \cite{das75} for a 3D electron gas. This RTA
formulation has not yet been extended beyond RPA. The effect of disorder
which leads to a damping of excitations enters the RPA dielectric function,
for a given electron-impurity collision frequency, through $\varepsilon _{%
\mathrm{RPA}}(k,\omega +i\gamma )$, where $\gamma $ is used as a model
parameter. For a degenerate electron gas (DEG) and for a given electron
density, the damping parameter can be assumed to be a constant to a good
approximation. The disorder-inclusive dielectric function, with the
collision frequency as a free parameter, allows some physical insight and
useful numerical estimates of the influence of disorder on energy loss in a
DEG. In 3D the predicted effect is a shorter life time with a smaller
propagation wavelength of plasmons resulting considerable modifications of
the SP (see, e.g., Refs.~\cite{ners02,ner03,ner04,ner08,ash80} and
references therein). For the stopping of a single ion, the broadening of the
plasmon peak with increasing $\gamma $ shifts the threshold for energy loss
by plasmon excitation towards lower projectile velocities. It now becomes
possible for low-velocity projectile ions to excite plasmons (in addition to
single-particle excitations). This increases the SP of 3D electron gas at
low projectile velocities, compared to the disorder-free RPA result \cite%
{ner04,ner08,ash80}. The situation with a 2D electron gas will be discussed
in detail in the following sections.

The second objective of our study is to investigate the influence of
exchange-correlation interaction (i.e. beyond RPA) in an electron gas on the
SP. For a 3D electron system it has been shown \cite{wan90,ma92} that the SP
in low and intermediate velocity regimes shows a definite increase due to
this interaction. A similar result has been reported for a 2D system \cite%
{wan95,wang95}. However let us note that if an asymptotic expansion of the
SP in a high-velocity regime is considered then it has been shown previously
that the first term in this expansion is unaffected by electron-electron
interaction. In this paper we calculate the next non-vanishing term of this
asymptotic expansion and show that it behaves as $B(r_{s})v^{-4}\ln
[A(r_{s})v]$, where exchange-correlation interactions are involved in $%
A(r_{s})$ and $B(r_{s})$. These functions depend on the target density
through Wigner-Seitz density parameter $r_{s}=(\pi n_{0}a_{0}^{2})^{-1/2}$,
where $n_{0}$ and $a_{0}$ are electron gas density and Bohr radius,
respectively. The details are presented in Sec.~\ref{sec:3}.

The plan of the paper is as follows. In Sec.~\ref{sec:2.1} we derive
analytical expressions for the disorder-inclusive dielectric function (DF)
for a 2D degenerate electron gas (DEG). We would like to mention that an
alternative but equivalent derivation is presented in Appendix~\ref{sec:app1}%
. The latter derivation contains certain attractive features. Through this
alternative formulation we consider a small-$k,\omega$ approximation for the
DF, and this approximate result is used in Sec.~\ref{sec:2.1}. In a small-$%
k,\omega$ approximation the plasmon dispersion for a disordered
two-dimensional DEG exhibits a constraint not present in 3D. This behavior
has been previously discussed in the literature \cite{giu84,gel86,gor91}. We
revisit this approximation through our formulation in Appendix~\ref{sec:app1}%
. The exact plasmon dispersion relations for an interacting DEG (including
exchange-correlation effects) are derived in Sec.~\ref{sec:2.2} by employing
local-field corrections to the RPA dielectric function. In Sec.~\ref{sec:3}
we briefly outline the general linear response function formalism of the 2D
stopping power of a point-like ion. After dealing with the excitation
equipartition in Sec.~\ref{sec:3.1} we develop, in Secs.~\ref{sec:3.2} and %
\ref{sec:3.3}, some analytical techniques to calculate the SP of an ion in
low- and high-velocity regimes. The two particular cases studied in these
sections are (i) low-velocity limit of the SP for an ion moving in a
disordered DEG, and (ii) high-velocity limit for a strongly interacting DEG.
Sec.~\ref{sec:4} contains systematic numerical calculations for the SP. The
results are summarized in Sec.~\ref{sec:sum} which also includes discussion
and outlook. Appendix~\ref{sec:app1} to which we draw the reader's attention
presents the above-mentioned alternative derivation of the DF for the
disorder-inclusive case in RPA, which is also valid in the complex $\omega$%
--plane. In Appendix~\ref{sec:app2} we provide some technical details for an
evaluation of the asymptotic SP.

\section{Dielectric function and dispersion relations for 2D electron gas}
\label{sec:2}

In the linear response theory, the stopping power (SP) of an external
projectile moving in a medium is related to the dielectric function $%
\varepsilon (k,\omega )$ of the medium. Both the single-particle and
collective excitations (i.e. the plasmons) contribute to the SP and these
contributions are contained in $\varepsilon (k,\omega )$ (see, e.g., Eq.~(%
\ref{eq:1}) below). In our study the two-dimensional (2D) target medium is
assumed to be disordered due to impurities etc. We shall incorporate effects
of disorder in $\varepsilon (k,\omega )$ in a somewhat phenomenological
manner. This is to include disorder through a relaxation time $\tau $ such
that the particle number is conserved. For a three-dimensional (3D) medium
this was done first by Mermin \cite{mer70} and then by Das \cite{das75} in
the random phase approximation (RPA) and in relaxation time approximation
(RTA). We refer the reader to \cite{mer70,das75} for details of this
formalism. For $\tau \rightarrow \infty $, this linear response function $%
\varepsilon (k,\omega ,1/\tau )$ reduces to the Lindhard dielectric function
\cite{lin54,lin64}. The dielectric function $\varepsilon (k,\omega )$ is
understood to contain $\gamma $ ($=1/\tau $) as a damping parameter due to
disorder. The form of $\varepsilon (k,\omega ,1/\tau )$ is to be specified
shortly for a 2D electron gas.

It is convenient to introduce the dimensionless Lindhard variables $%
z=k/2k_{F}$, $u=\omega /kv_{F}$, where $v_{F}$\ and $k_{F}$\ ($=(2\pi
n_{0})^{1/2}$) are, respectively, the Fermi velocity and wave number of the
target electrons. Also we introduce the density parameters $%
\chi^{2}=1/\alpha =1/k_{F}a_{0}=r_{s}/\sqrt{2}$. In our calculations $\chi $
and $\alpha $ (or $r_{s}$) serve as a measure of electron density. (Note
that the density parameter $\chi $ introduced above differs from usual
definition by a factor $\pi $ see, e.g., Refs.~\cite{bre93,bre94,bre98,wan95}).

\subsection{Disordered electron gas: RPA}
\label{sec:2.1}

Let us now specify the disorder-inclusive dielectric function for 2D
zero-temperature (degenerate) electron gas (DEG). This has been done
previously in Refs.~\cite{giu84,gel86,gor91} employing small-$k,\omega $
approximation. Here within RPA and RTA we derive the disorder-inclusive
dielectric function (DF) without further approximations on the
energy-momentum spectrum i.e. on $\omega $ and $k$. As pointed out in Ref.~%
\cite{gel86} the physical arguments for deriving number-conserving DF by
Mermin \cite{mer70} and Das \cite{das75} in 3D are independent of
dimensionality. Therefore with the notations introduced in the preceding
paragraph, the DF for 2D DEG reads
\begin{equation}
\varepsilon (z,u)=1+\frac{(zu+i\Gamma )\left[ \varepsilon _{\mathrm{RPA}%
}(z,u,\Gamma )-1\right] }{zu+i\Gamma \left[ \varepsilon _{\mathrm{RPA}%
}(z,u,\Gamma )-1\right] /\left[ \varepsilon _{\mathrm{RPA}}(z,0)-1\right] },
\label{eq:2}
\end{equation}%
where $\Gamma =\hbar \gamma /4E_{F}$, $E_{F}$ being the Fermi energy $=\hbar
^{2}k_{F}^{2}/2m$ with $m$ as the effective mass. The quantity $\gamma $ (or
$\Gamma $) is a measure of damping of excitations in the disordered electron
gas. $\varepsilon _{\mathrm{RPA}}(z,u,\Gamma )=\varepsilon _{\mathrm{RPA}%
}(k,\omega +i\gamma )$ is the longitudinal dielectric function of DEG in the
RPA derived in 2D by Stern \cite{ste67}. $\varepsilon _{\mathrm{RPA}%
}(z,0)=\varepsilon _{\mathrm{RPA}}(k,0)$ is the static dielectric function.
We have analytically evaluated the disorder-inclusive $\varepsilon (z,u)$
for which the results, presented below, appear to be new and we have
utilized them in our numerical investigation.

Let us recall the Lindhard (RPA) expression for the longitudinal dielectric
function \cite{lin54}. In variables $z$ and $u$ and in 2D it reads as \cite%
{ste67}
\begin{eqnarray}
\varepsilon _{\mathrm{RPA}}\left( z,u,\Gamma \right) &=&1+\frac{\chi ^{2}z}{%
\pi }\int_{0}^{\pi }d\theta \int_{0}^{1}\frac{qdq}{z^{4}-\left( uz+i\Gamma
+qz\cos \theta \right) ^{2}}  \label{eq:3} \\
&=&1+\frac{\chi ^{2}}{2z^{2}}\left[ F_{1}(z,u,\Gamma )+iF_{2}(z,u,\Gamma )%
\right] ,  \nonumber
\end{eqnarray}%
where we have split explicitly the DF $\varepsilon _{\mathrm{RPA}}\left(
z,u,\Gamma \right) $ into the real and imaginary parts and have introduced
the real functions (for real $z$ and $u$) $F_{1}(z,u,\Gamma )$ and $%
F_{2}(z,u,\Gamma )$ as in the usual RPA expression of longitudinal
dielectric function.

Performing the $q$ and $\theta $ integrations in Eq.~(\ref{eq:3}) \cite%
{gra80} we obtain, for a non-zero damping,
\begin{eqnarray}
F_{1}(z,u,\Gamma ) &=&2z+\frac{\Gamma }{z}\left[ Y_{-}\left( z,u_{-}\right)
-Y_{-}\left( z,u_{+}\right) \right]  \label{eq:4} \\
&&+\left( u_{-}-1\right) Y_{+}\left( z,u_{-}\right) -\left( u_{+}-1\right)
Y_{+}\left( z,u_{+}\right) ,  \nonumber
\end{eqnarray}%
\begin{eqnarray}
F_{2}(z,u,\Gamma ) &=&\frac{\Gamma }{z}\left[ Y_{+}\left( z,u_{-}\right)
-Y_{+}\left( z,u_{+}\right) \right]  \label{eq:5} \\
&&+\left( u_{+}-1\right) Y_{-}\left( z,u_{+}\right) -\left( u_{-}-1\right)
Y_{-}\left( z,u_{-}\right)  \nonumber
\end{eqnarray}%
with $u_{\pm }=u\pm z$,
\begin{equation}
Y_{\pm }\left( z,u\right) =\frac{1}{\sqrt{2}}\sqrt{\sqrt{\frac{z^{2}\left(
u+1\right) ^{2}+\Gamma ^{2}}{z^{2}\left( u-1\right) ^{2}+\Gamma ^{2}}}\pm
\frac{z^{2}\left( u^{2}-1\right) +\Gamma ^{2}}{z^{2}\left( u-1\right)
^{2}+\Gamma ^{2}}} .
\label{eq:6}
\end{equation}%
In the case of vanishing damping ($\gamma \rightarrow 0$ and $\Gamma
\rightarrow 0$) the expressions (\ref{eq:2})-(\ref{eq:6}) coincide with the
Stern result \cite{ste67} with
\begin{equation}
f_{1}(z,u)=\left. F_{1}(z,u,\Gamma )\right\vert _{\Gamma \rightarrow
0}=2z+C_{-}\sqrt{u_{-}^{2}-1}-C_{+}\sqrt{u_{+}^{2}-1} ,
\label{eq:7}
\end{equation}%
\begin{equation}
f_{2}(z,u)=\left. F_{2}(z,u,\Gamma )\right\vert _{\Gamma \rightarrow 0}=D_{-}%
\sqrt{1-u_{-}^{2}}-D_{+}\sqrt{1-u_{+}^{2}} ,
\label{eq:8}
\end{equation}

\begin{equation}
D_{\pm }=H\left( 1-\left\vert u_{\pm }\right\vert \right) ,\qquad C_{\pm
}=H\left( \left\vert u_{\pm }\right\vert -1\right) \frac{u_{\pm }}{%
\left\vert u_{\pm }\right\vert } .
\label{eq:9}
\end{equation}%
Here $H(z)$ is the Heaviside unit-step function. The static DF involved in
Eq.~(\ref{eq:2}) can be found either from Eqs.~(\ref{eq:4}) and (\ref{eq:5})
at the limits $u\rightarrow 0$, $\Gamma \rightarrow 0$ or from Eqs.~(\ref%
{eq:7})-(\ref{eq:9}) at $u\rightarrow 0$. The result reads
\begin{equation}
\varepsilon _{\mathrm{RPA}}\left( z,0\right) =1+\frac{\chi ^{2}}{z^{2}}%
f\left( z\right)
\label{eq:10}
\end{equation}%
with
\begin{equation}
f\left( z\right) =\frac{1}{2}f_{1}(z,0)=\left\{
\begin{array}{cc}
z, & 0\leqslant z\leqslant 1 \\
\frac{1}{z+\sqrt{z^{2}-1}}, & z>1%
\end{array}%
\right. .
\label{eq:11}
\end{equation}

\begin{figure}[tbp]
\includegraphics[width=14cm]{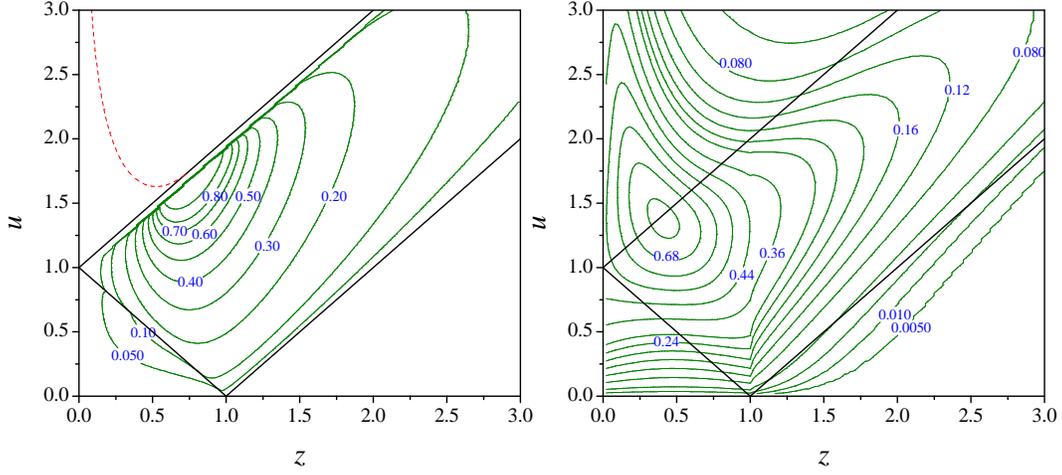}
\caption{(Color online) The contour plot of $L(z,u)=\mathrm{Im}[-1/\protect%
\varepsilon (z,u)]$ as a function of the variables $z$ and $u$ for $r_{s}=2$
and without (left panel) and with (right panel) damping ($\hbar \protect%
\gamma =15$ eV). The dashed line in left panel shows the plasmon dispersion
curve $u_{r}(z)$ (see the text for details) with $\protect\gamma =0$. The
numbers indicate the values of $L(z,u)$.}
\label{fig:1}
\end{figure}

To demonstrate the effect of the damping in Fig.~\ref{fig:1} we show the
contour plots of the energy loss function $L(z,u)=\mathrm{Im}[-1/\varepsilon
(z,u)]$ without (left panel) and with (right panel) damping. The plasmon
dispersion function $u_{r}\left( z\right) $\ in the left panel is also shown
as a dashed line (the explicit derivation of the plasmon dispersion curve $%
u_{r}\left( z\right) $ without damping is given below in Sec.~\ref{sec:2.2},
see Eqs.~(\ref{eq:17}), (\ref{eq:20}) and Fig.~\ref{fig:3}). The
single-particle excitations energies $\hbar \omega _{\mathrm{sp}}=\left\vert
\hbar kv_{F}\pm \hbar ^{2}k^{2}/2m\right\vert $ (or $u=\left\vert z\pm
1\right\vert $ in dimensionless units) are demonstrated as thick solid
lines. As expected the energy loss function $L(z,u)$ in the case of
vanishing damping (left panel) is localized in the domains $0<u<1-z$
with $0<z<1$, $\vert z-1\vert <u<z+1$ as well as on the plasmon curve $u_{r}\left( z\right)
$\ where the function $L(z,u)$ behaves as a Dirac $\delta $-function and
becomes infinity. In the case of non-zero damping (right panel) the energy
loss function is broadened due to the damping and becomes non-zero also in
the domains $u<\left\vert z-1\right\vert $ and $u>z+1$.

\begin{figure}[tbp]
\includegraphics[width=14cm]{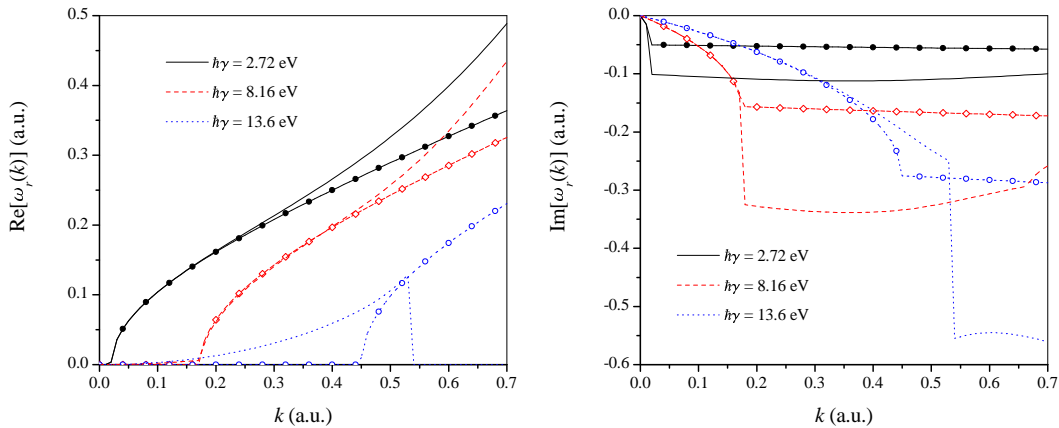}
\caption{(Color online) Real ($\mathrm{Re}[\protect\omega _{r}(k)]$, left
panel) and imaginary ($\mathrm{Im}[\protect\omega _{r}(k)]$, right panel)
parts (in a.u.) of exact (the lines without symbols) and approximate (the
lines with symbols) solutions of dispersion equation vs $k$ (in a.u.) for $%
k\leqslant 2k_{F}$, $r_{s}=4$, $\hbar \protect\gamma =2.72$ eV (solid
lines), $\hbar \protect\gamma =8.16$ eV (dashed lines), $\hbar \protect%
\gamma =13.6$ eV (dotted lines).}
\label{fig:2}
\end{figure}

Equations (\ref{eq:2})-(\ref{eq:5}) constitute the number-conserving DF for
a 2D disordered electron gas. Deriving these expressions we have explicitly
split the DF into real and imaginary parts assuming real variables $z$ and $u
$. An alternative (but equivalent) expression for this DF is derived in
Appendix~\ref{sec:app1} which is valid for any complex $\omega $ and $k$.
With this exact (within RPA and RTA) expression in Appendix~\ref{sec:app1}
we then calculate the DF within small $\omega ,k$-approximation obtained
previously in Refs.~\cite{giu84,gel86,gor91} and revisited in Appendix~\ref%
{sec:app1}. The basic feature of this approximation is the prediction of the
threshold condition for plasmon propagation which is absent in 3D (see,
e.g., Ref.~\cite{ner04}). Indeed the solution of the dispersion equation $%
\varepsilon \left( k,\omega \right) =0$, where $\varepsilon \left( k,\omega
\right) $ is given by Eq.~(\ref{eq:ap8}), reads \cite{giu84,gel86,gor91}
\begin{equation}
\omega _{r}\left( k\right) =\frac{1+k\lambda _{\mathrm{TF}}}{1+k\lambda _{%
\mathrm{TF}}/2}\left[ -\frac{i\gamma }{2}+\sqrt{\omega _{p}^{2}\left(
k\right) \left( 1+\frac{k\lambda _{\mathrm{TF}}}{2}\right) -\frac{\gamma ^{2}%
}{4}}\right] ,
\label{eq:12}
\end{equation}%
where $\omega _{p}^{2}\left( k\right) =2\pi n_{0}e^{2}k/m$ is the plasma
frequency for a 2D electron gas. The condition that $\omega _{r}\left(
k\right) $ has a real part (for plasmon propagation) leads to $k>k_{\ast }$,
where
\begin{equation}
k_{\ast }=k_{\mathrm{TF}}\left[ \sqrt{1+\left( \frac{\gamma }{k_{\mathrm{TF}%
}v_{F}}\right) ^{2}}-1\right]
\label{eq:13}
\end{equation}%
with $k_{\mathrm{TF}}=1/\lambda _{\mathrm{TF}}=2/a_{0}$. Thus, within small $%
\omega ,k$-approximation, disorder in 2D electronic systems considerably
softens plasmons; they cannot propagate for $k<k_{\ast }$ and their
dispersion relation is strongly altered relative to the collisionless case.
However, since these results were obtained in small $\omega ,k$--domain one
can expect some modifications for large momentum transfers at $k\gtrsim
k_{\ast }$. Figure~\ref{fig:2} shows the real (left panel) and imaginary
(right panel) parts of the solutions of the dispersion equations with
approximate (Eq.~(\ref{eq:12})) and exact dielectric functions, Eqs.~(\ref%
{eq:ap3})-(\ref{eq:ap6})). For simplicity we consider the case $k\leqslant
2k_{F}$ when the function $Q$ in Eq.~(\ref{eq:ap5}) vanishes. Note that the
condition $k>k_{\ast }$ together with the inequality $k\leqslant 2k_{F}$
requires that $\hbar \gamma /E_{\mathrm{H}}<4r_{s}^{-2}\sqrt{1.414r_{s}+1}$
with $E_{\mathrm{H}}=me^{4}/\hbar ^{2}\simeq 27.2$ eV. It is seen that the
slope of the imaginary part of $\omega _{r}\left( k\right) $\ (right panel)
is dramatically changed at some value of $k$ where the expression under
square root in Eq.~(\ref{eq:12}) changes the sign. For small $\omega ,k$%
-approximation this value of $k$ is given by Eq.~(\ref{eq:13}). As pointed
out in Appendix~\ref{sec:app1} the approximation (\ref{eq:12}) is valid when
one neglects the single-particle energy $\hbar \omega _{k}=\hbar
^{2}k^{2}/2m $ with respect to $\hbar kv_{F}$. Therefore, in general, we
expect good agreement between approximate and exact $\omega _{r}\left(
k\right) $\ for small momentum $k$, as shown in Fig.~\ref{fig:2}. However,
with increasing $\gamma $ the approximate dispersion relation (\ref{eq:12})
fails to predict $\omega _{r}\left( k\right) $ correctly. As shown in Fig.~%
\ref{fig:2} (left panel, dotted curve with $\hbar \gamma =13.6$ eV) at small
$k$ ($k\lesssim k_{\ast }$) the energy of plasmons $\omega _{r}\left(
k\right) $ is not exactly zero as predicted by Eq.~(\ref{eq:12}) although
the probability of plasmon generation is strongly reduced due to the
relation $\mathrm{Re}(\omega _{r})\ll \mathrm{Im}(\omega _{r})$. Moreover,
in contrast to the predictions of approximation (\ref{eq:12}) in this case
with increasing momentum $k$ the real part of $\omega _{r}$ vanishes and
plasmons cannot propagate any more.

\subsection{Strongly coupled electron gas: Beyond RPA}
\label{sec:2.2}

In this section we consider exchange-correlation interaction effects via
local-field corrected (LFC) DF but we neglect the disorder (i.e. $\gamma =0$%
). To include disorder in a fully interacting electron gas at a microscopic
level is rather involved, and no analytical calculations of $\varepsilon
(k,\omega )$ without restrictions on $k$ and $\omega $ are still available.
An attempt to involve strong correlations in RTA and within small $k,\omega $%
-approximation (see Eq.~(\ref{eq:12})) has been done in Ref.~\cite{gel86}.
Instead we employ here the LFC dielectric function and demonstrate some
useful results which have not been considered previously. Our discussion
below is based on the LFC dielectric function of a fully DEG see, e.g., Ref.~%
\cite{wan95} (in dimensionless variables $z$ and $u$)
\begin{equation}
\varepsilon (z,u)=1+\frac{\mathcal{P}(z,u)}{1-G(z)\mathcal{P}(z,u)},
\label{eq:14}
\end{equation}%
where $\mathcal{P}(z,u)$\ is the polarizability of the free-electron gas
obtained in RPA by Stern \cite{ste67}
\begin{equation}
\mathcal{P}(z,u)=\varepsilon _{\mathrm{RPA}}\left( z,u\right) -1=\frac{\chi
^{2}}{2z^{2}}\left[ f_{1}(z,u)+if_{2}(z,u)\right]
\label{eq:15}
\end{equation}%
with $\varepsilon _{\mathrm{RPA}}\left( z,u\right) =\varepsilon _{\mathrm{RPA%
}}\left( z,u,\Gamma \rightarrow 0\right) $, where $\varepsilon _{\mathrm{RPA}%
}\left( z,u,\Gamma \right) $, $f_{1}(z,u)$ and $f_{2}(z,u)$ are given by
Eqs.~(\ref{eq:3}), (\ref{eq:7}) and (\ref{eq:8}), respectively. Note that
our definition of the functions $f_{1}(z,u)$ and $f_{2}(z,u)$ differs from
the definition given in Refs.~\cite{wan95,wang95} by a factor of $-1/2$. $%
G(z)$ is the LFC function, which includes the effects of
exchange-correlation interactions. Within a sum-rule version of the
self-consistent approach, Gold and Calmels presented \cite{gol93} a
parameterized expression $G(z)$ for the 2D electron gas,
\begin{equation}
G(z)=\frac{zG_{0}\left( r_{s}\right) }{\sqrt{G_{12}^{2}\left( r_{s}\right)
+z^{2}G_{22}^{2}\left( r_{s}\right) }} .
\label{eq:16}
\end{equation}%
The coefficients $G_{0}(r_{s})$, $G_{12}(r_{s})$ and $G_{22}(r_{s})$ are
determined by $G_{0}(r_{s})=1.983r_{s}^{1/3}$, $%
G_{12}(r_{s})=1.626C_{12}(r_{s})$, $G_{22}(r_{s})=\sqrt{2}%
r_{s}^{-1/3}C_{22}(r_{s})$, with $C_{12}(r_{s})=\alpha _{1}r_{s}^{\gamma
_{1}}$, $C_{22}(r_{s})=\alpha _{2}r_{s}^{\gamma _{2}}$, and the parameters $%
\alpha _{1}$, $\alpha _{2}$ and $\gamma _{1}$, $\gamma _{2}$\ can be found
in Ref.~\cite{gol93}.

Now we consider the exact solution of the dispersion equation $\varepsilon
(z,u)=0$ for an interacting electron gas when the DF is given by LFC
expression (\ref{eq:14}). From Eqs.~(\ref{eq:7})-(\ref{eq:9}) and (\ref%
{eq:14}), (\ref{eq:15}) it is seen that the collective plasma modes
(plasmons) can propagate with the frequency and momentum $\omega $ and $k$
(or $u$ and $z$) which lie in the domain $u\geqslant z+1$ where $%
f_{2}(z,u)=0 $ and $\mathrm{Im}[\varepsilon (z,u)]=0$. In this domain the
dispersion equation has an exact analytical solution which, in Lindhard's
dimensionless variables, is given by
\begin{equation}
u_{r}^{2}\left( z\right) =1+z^{2}\left[ \alpha zg\left( z\right) +1\right]
^{2}+\frac{1}{\alpha zg\left( z\right) \left[ \alpha zg\left( z\right) +2%
\right] }
\label{eq:17}
\end{equation}%
with $\alpha =\sqrt{2}/r_{s}$ and $g\left( z\right) =[1-G(z)]^{-1}$. It is
straightforward to check that the solution (\ref{eq:17}) indeed satisfies
the condition $u_{r}(z)\geqslant z+1$ for arbitrary $z$. However, an
inspection of the dispersion equation shows that this solution exists only
for the wave numbers from the domain $0\leqslant k\leqslant k_{c}$ (or $%
0\leqslant z\leqslant z_{c}$) where the critical wave number $z_{c}$ is
obtained from an equation $u_{r}(z_{c})=1+z_{c}$, i.e. in this point the
plasmon curve $u_{r}(z)$ touches to the boundary of the single-particle
continuum $u=1+z$. Explicitly, the critical wave numbers are determined from
transcendental equation
\begin{equation}
\alpha z_{c}^{2}g\left( z_{c}\right) \left[ 2+\alpha z_{c}g\left(
z_{c}\right) \right] =1 .
\label{eq:18}
\end{equation}%
Table~\ref{tab:1} shows the quantity $z_{c}$ and the minimum of the
dispersion function $\lambda _{c}=u_{r}(z_{\min })$ with $u_{r}^{\prime
}(z_{\min })=0$\ for some values of the density parameter $r_{s}$. The
critical wave numbers and the quantities $\lambda _{c}$ (labeled as $%
z_{c}^{(0)}$ and $\lambda _{c}^{(0)}$, respectively) are also shown for
non-interacting electron gas, i.e. with $G(z)=0$ and $g(z)=1$. These
quantities are important for evaluation of the SP in Sec.~\ref{sec:3}.

\begin{table}[tbp]
\caption{The critical wave numbers (dimensionless) $z_{c}$ and the minimum
values $\protect\lambda _{c}=u_{r}(z_{\min })$ of the dispersion function $%
u_{r}(z)$ for some values of the density parameter $r_{s}$. $z_{c}^{(0)}$
and $\protect\lambda _{c}^{(0)}$ represent the smae quantities but for
non-interacting 2D electron gas.}
\label{tab:1}
\begin{center}
\begin{tabular}{p{1.1cm}p{1.1cm}p{1.1cm}p{1.1cm}p{1.1cm}p{1.1cm}p{1.1cm}p{1.1cm}p{1.1cm}p{1.1cm}}
\hline\hline
$r_{s}$ & 0.10 & 0.50 & 1.00 & 1.50 & 2.00 & 2.50 & 3.00 & 3.50 & 4.00 \\
\hline
$z_{c}$ & 0.126 & 0.288 & 0.390 & 0.457 & 0.507 & 0.546 & 0.579 & 0.606 &
0.629 \\
$z_{c}^{(0)}$ & 0.135 & 0.345 & 0.510 & 0.638 & 0.748 & 0.845 & 0.932 & 1.013
& 1.089 \\
$\lambda_{c}$ & 1.116 & 1.264 & 1.358 & 1.421 & 1.469 & 1.507 & 1.538 & 1.565
& 1.588 \\
$\lambda^{(0)}_{c}$ & 1.122 & 1.304 & 1.440 & 1.543 & 1.629 & 1.704 & 1.770
& 1.831 & 1.886 \\ \hline
\end{tabular}%
\end{center}
\end{table}

We can present the dispersion expression~(\ref{eq:17}) obtained above, in
the usual form
\begin{eqnarray}
\omega _{r}^{2}\left( k\right) &=&\omega _{p}^{2}\left( k\right) \frac{%
2/g\left( k\right) +k\lambda _{\mathrm{TF}}g\left( k\right) }{2+k\lambda _{%
\mathrm{TF}}g\left( k\right) }  \label{eq:19} \\
&&+k^{2}v_{F}^{2}\frac{2-\frac{1}{2}g\left( k\right) +k\lambda _{\mathrm{TF}%
}g\left( k\right) }{2+k\lambda _{\mathrm{TF}}g\left( k\right) }+\frac{\hbar
^{2}k^{4}}{4m^{2}}\left[ 1+k\lambda _{\mathrm{TF}}g\left( k\right) \right]
^{2}  \nonumber
\end{eqnarray}%
which for vanishing exchange-correlation interactions (i.e. at $g\left(
k\right) =1$) reads
\begin{equation}
\omega _{r}^{2}\left( k\right) =\omega _{p}^{2}\left( k\right)
+k^{2}v_{F}^{2}\frac{3+2k\lambda _{\mathrm{TF}}}{4+2k\lambda _{\mathrm{TF}}}+%
\frac{\hbar ^{2}k^{4}}{4m^{2}}\left( 1+k\lambda _{\mathrm{TF}}\right) ^{2}.
\label{eq:20}
\end{equation}%
This exact (within the employed model) dispersion relation may be compared
with an approximate result derived by Fetter within a hydrodynamical
approach \cite{fet73}. Equation~(\ref{eq:20}) agrees with the hydrodynamic
result if the last term (the single-particle energy) in this expression is
neglected and the coefficient at $k^{2}v_{F}^{2}$ is replaced by a constant
factor $1/2$. It should be emphasized that in general and at long wavelengths
$\omega _{r}(k)$ from Eq.~(\ref{eq:19}) for an interacting 2D
electron system varies like $k^{1/2}$ independently of the LFC $G(k)$ and in
contrast to the 3D case. This latter behavior seems first to have been
suggested by Ferrell \cite{fer58} and later investigated in more detail by
Stern \cite{ste67} (see also the review paper \cite{and82}). It arises from
the electromagnetic fields in the vacuum surrounding the plane, with an
associated reduction in the screening. Since $\omega _{r}(k)$ increases
monotonically from zero, an external perturbation of arbitrarily low
frequency can always excite collective modes. Hence, the characteristic 3D
absorption edge at constant 3D $\omega _{p}$ is here entirely absent.
Moreover, the group and phase velocities both diverge like $k^{-1/2}$ as $%
k\rightarrow 0$.

\begin{figure}[tbp]
\includegraphics[width=8cm]{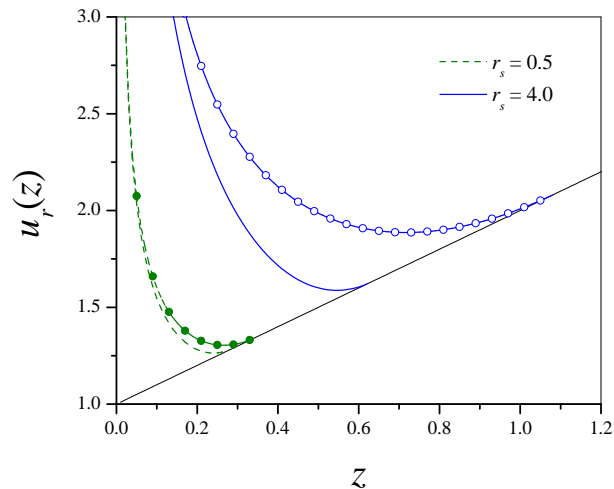}
\caption{(Color online) The dispersion curve $u_{r}(z)$ for free (lines with
symbols) and interacting (lines without symbols) electron gas vs $z$ for $%
r_{s}=0.5$ (dashed lines) and $r_{s}=4.0$ (solid lines). The oblique solid
line corresponds to $\protect\omega =kv_{F}+\hbar k^{2}/2m$ (or $u=z+1$).}
\label{fig:3}
\end{figure}

Figure~\ref{fig:3} shows the plasmon dispersion curve $u_{r}(z)$ for
interacting (the lines without symbols) and non-interacting (the lines with
symbols) electron gas, i.e. Eqs.~(\ref{eq:17}) and (\ref{eq:20}),
respectively. The points where the plasmon curves touch the single-particle
excitation boundary are given by $z_{c}$ or $z_{c}^{(0)}$, see Table~\ref%
{tab:1}. It is seen that the exchange-correlation interaction may strongly
reduce the values of $u_{r}(z)$. It must be pointed out a technical but
important detail which, to our knowledge, has not been yet discussed in the
literature. From Fig.~\ref{fig:3} it is seen that in $u,z$ plane the plasmon
curve $u_{r}(z)$ has a minimum which is absent in usual units $\omega ,k$
where $\omega _{r}(k)$ is a monotonic increasing function. By interchanging
the $z$ and $u$ axes in Fig.~\ref{fig:3} one obtains the plasmon dispersion
curve $z_{r}(u)$ which, however, in contrast to the 3D case has two
different branches with increasing ($z_{r1}(u)$) and decreasing ($z_{r2}(u)$%
) dispersion functions (at the minimum of $u_{r}(z)$ both $z_{r1}(u)$ and $%
z_{r2}(u)$ curves contact each other). The dispersion relations $z_{r1}(u)$
and $z_{r2}(u)$ can be provisionally treated as the "single-particle" and
"plasmonic" relations, respectively. Therefore when one attempts to perform $%
z$-integration in Eq.~(\ref{eq:1}) before $u$-integration, as was done in
Ref.~\cite{wan95}, the double integration in the SP is reduced to two line
integrations along the contours $z_{r1}(u)$ and $z_{r2}(u)$ and both of them
contribute to the SP. In other words in this case the energy loss function $%
L(z,u)$ introduced above contains two Dirac $\delta $-functions. In fact, we
see from our numerical calculations that near the SP maximum the
contribution of $z_{r1}(u)$ is not necessarily small compared to the
contribution of the other one, $z_{r2}(u)$. Although the total contributions
of both in the SP are in general much smaller than the purely
single-particle contributions. This is a violation of the Lindhard-Winther
equipartition sum rule \cite{lin64} which we further discuss in Sec.~\ref%
{sec:3.1}. To avoid this technical problem in the numerical calculations it
is easier to perform first in Eq.~(\ref{eq:1}) the $u$-integration and then
using the dispersion function $u_{r}(z)$ given by Eq.~(\ref{eq:17}).

\section{Stopping power}
\label{sec:3}

With the theoretical formalism presented so far, we now take up the main
topic of this paper. This is to study the stopping power (SP) of a
point-like ion in a 2D degenerate electron gas as well as to show how
collective and single-particle excitations in the target medium DEG
contribute to the SP. And, as in the previous section, we shall present new
theoretical results within the linear response approach. We consider two
models for a DEG in 2D. (i) A disordered DEG for which we use a
number-conserving DF given in Eqs.~(\ref{eq:2})-(\ref{eq:6}). For this case
we present analytical calculations and new results for the SP in a
low-velocity limit. (ii) A strongly coupled DEG with a DF which includes
LFC, Eqs.~(\ref{eq:14}) and (\ref{eq:15}). This case has been studied in
Refs.~\cite{wan95,wang95} where the leading term in a high-velocity limit of
the SP is calculated using a plasmon-pole approximation. This calculation is
supported by a more rigorous treatment, again for the leading term only, in
Ref.~\cite{bal06} which is based on a method of moments and includes
electron-electron interactions. Now the leading term happens not to depend
on electron-electron interaction. It is then of interest to calculate
analytically the next non-vanishing terms of the high-velocity SP. As shown
below these terms are significantly modified by electron-electron
interaction and thus are more involved than the leading term.

We consider an external point-like projectile of carge $Ze$ moving with
velocity $\mathbf{v}$ in a homogeneous and isotropic 2D electron medium
characterized by the dielectric function $\varepsilon (k,\omega )$ or $%
\varepsilon (z,u)$. Then in the linear response theory the SP which is the
energy loss per unit length by this projectile is given by \cite{bre93,wan95}
\begin{equation}
S=\frac{8\Sigma _{0}Z^{2}}{\pi \chi ^{4}\lambda }\int_{0}^{\lambda }\frac{udu%
}{\sqrt{\lambda ^{2}-u^{2}}}\int_{0}^{\infty }\mathrm{Im}\frac{-1}{%
\varepsilon \left( z,u\right) }zdz .
\label{eq:1}
\end{equation}%
Here $\lambda =v/v_{F}$, $\Sigma _{0}=e^{2}/a_{0}^{2}\simeq 5.132$ GeV/cm $%
=51.32$ eV/\AA . We have used the Lindhard variables $z$ and $u$ introduced
in Sec.~\ref{sec:2}. In our calculations we shall consider the range of $v$
for which the linear response theory is found to be adequate \cite{zwi99}.

\subsection{Equipartition sum rule}
\label{sec:3.1}

With the theoretical formalism presented so far, we now take up one of the
main topic of this paper. This is to study how collective and
single-particle excitations in the 2D electron gas contribute to the SP.
This problem was first addressed by Lindhard and Winther \cite{lin64} (LW)
for a 3D degenerate electron gas without damping ($\gamma =0$). They
formulated an equipartition sum rule which states that an integral
proportional to that in Eq.~(\ref{eq:1})
\begin{equation}
\Im \left( u\right) =\Im _{sp}\left( u\right) +\Im _{p}\left( u\right)
=\int_{0}^{\infty }\mathrm{Im}\frac{-1}{\varepsilon \left( z,u\right) }zdz
\label{eq:21}
\end{equation}%
receives equal contributions from plasmon ($\Im _{p}$) (with $0<z<u-1$) and
from single-particle excitations ($\Im _{sp}$) (with $u-1<z<u+1$),
respectively. The functions $\Im _{p}(u)$ and $\Im _{sp}(u)$ may then be
written as
\begin{equation}
\Im _{p}\left( u\right) =\int_{0}^{u-1}\mathrm{Im}\frac{-1}{\varepsilon
\left( z,u\right) }zdz=\frac{\pi z_{r}\left( u\right) }{\left\vert \frac{%
\partial }{\partial z}\varepsilon \left( z,u\right) \right\vert
_{z=z_{r}\left( u\right) }} ,
\label{eq:22}
\end{equation}

\begin{equation}
\Im _{sp}\left( u\right) =\int_{u-1}^{u+1}\mathrm{Im}\frac{-1}{\varepsilon
\left( z,u\right) }zdz .
\label{eq:23}
\end{equation}%
Here $z_{r}\left( u\right) $\ is the solution of the dispersion equation $%
\varepsilon (z,u)=0$ (the inverse of the dispersion function $u_{r}\left(
z\right) $). This equipartition rule is valid for sufficiently large $u$, $%
u>u_{m}$, where the threshold value $u_{m}$ in 3D case is obtained from the
equation $z_{r}\left( u_{m}\right) =u_{m}-1$. In recent works \cite%
{ners02,ner08,ner02} we have shown that the LW equipartition rule does not
necessarily hold for an extended charged projectile e.g. a diproton cluster
in a 3D degenerate electron gas without disorder ($\gamma =0$) as well as
for a point-like ion in a disordered DEG. We have established some
generalized stopping power sum rules. In this section we briefly show that
the LW equipartition rule is also violated for a 2D electron gas. In the
present context it should be emphasized that the plasmon contribution given
by Eq.~(\ref{eq:22}) contains indeed two terms, with $z_{r1}\left( u\right) $
and $z_{r2}\left( u\right) $, as discussed above. The existence of both
branches requires the threshold condition $u>u_{m}$, where $u_{m}$\ is the
minimum value of the dispersion function $u_{r}\left( z\right) $ shown, e.g.
in Fig.~\ref{fig:3}. However, it is clear that the contribution of $%
z_{r1}\left( u\right) $ vanishes at $u>u_{c}$, where $%
u_{c}=u_{r}(z_{c})>u_{m}$\ (the point where the plasmon curve touches to the
single-particle excitations boundary). For simplicity we consider below only
the domain $u>u_{c}$ where only $z_{r2}\left( u\right) \equiv z_{r}\left(
u\right) $ contributes to the SP integral (\ref{eq:22}). As an example we
employ the DF (\ref{eq:14}) together with Eqs.~(\ref{eq:7})-(\ref{eq:9}) and
(\ref{eq:15}) for an interacting DEG. The simplest way to show the violation
of the LW equipartition rule in 2D is to calculate the asymptotic values of
the contributions $\Im _{p}\left( u\right) $\ and $\Im _{sp}\left( u\right) $
at $u\gg 1$. The inverse dispersion function $z_{r}\left( u\right) $ for 2D
interacting DEG is evaluated in Appendix~\ref{sec:app2}, see Eqs.~(\ref%
{eq:app1})-(\ref{eq:app2}). Using these expressions it is straightforward to
calculate the single-particle and collective contributions to the SP
integral which at $u\gg 1$\ become
\begin{eqnarray}
\Im _{p}\left( u\right) &\simeq &\frac{\pi }{4\alpha ^{2}u^{4}}\left\{ 1+%
\frac{3-2\varkappa \left( r_{s}\right) }{2u^{2}}+\frac{3}{4u^{4}}\left[
\varkappa ^{2}\left( r_{s}\right) -3\varkappa \left( r_{s}\right) +\frac{29}{%
12}\right] +...\right\} ,  \label{eq:24} \\
\Im _{sp}\left( u\right) &\simeq &\frac{\pi }{4\alpha u}\left\{ 1+\frac{1}{%
4u^{2}}+\frac{1}{2\alpha u^{3}}\left[ 1-\frac{\varkappa \left( r_{s}\right)
}{\varkappa _{0}\left( r_{s}\right) }\right] +\frac{1}{8u^{4}}+...\right\} .
\label{eq:25}
\end{eqnarray}%
Here $\varkappa _{0}\left( r_{s}\right) $ and $\varkappa \left( r_{s}\right)
$ are defined in Appendix~\ref{sec:app2}. From the above expressions it is
clear that the contribution of the collective excitations is much smaller
than the contribution from single-particle excitations, $\Im _{p}\left(
u\right) \ll \Im _{sp}\left( u\right) $, which indicates the violation of
the LW equipartition rule. A similar result has been found numerically in
Ref.~\cite{wan95} and is supported by our own numerical calculations. Of
course, Eqs.~(\ref{eq:24}) and (\ref{eq:25}) are not strong results. An
exact treatment can be developed on the basis of the integration contour on
the complex $z$-plane suggested by LW \cite{lin64} and investigated in
details in Ref.~\cite{ner08}. The technique developed in \cite{ner08} is
independent of the dimensionality of electron gas but requires a necessary
analytic continuation of the DF in the complex $z$-plane, that is $%
\varepsilon (-z^{\ast },u)=\varepsilon ^{\ast }(z,u)$, where the asterix
indicates a complex conjugate quantity. It is easy to see that this
condition is violated for a 2D electron gas. For simplicity let us consider
non-interacting DEG with the DF given by Eq.~(\ref{eq:3}) in the integral
form and with $\Gamma \rightarrow +0$. In this case one can easily check
that $\varepsilon _{\mathrm{RPA}}\left( -z^{\ast },u\right) =2-\varepsilon _{%
\mathrm{RPA}}^{\ast }\left( z,u\right) $ (a similar equation can be obtained
for an interacting electron gas). Therefore an arbitrary function of the
form
\begin{equation}
\varepsilon _{\mathrm{eff}}\left( z,u\right) =1+\frac{\mathcal{C}}{z}\left[
\varepsilon _{\mathrm{RPA}}\left( z,u\right) -1\right]
\label{eq:26}
\end{equation}%
with an arbitrary constant $\mathcal{C}$ defines an effective DF of a 2D
electron gas which satisfies the required condition, i.e. $\varepsilon _{%
\mathrm{eff}}(-z^{\ast },u)=\varepsilon _{\mathrm{eff}}^{\ast }(z,u)$.
Applying now the contour integration technique developed in Ref.~\cite{ner08}
one can strongly prove that the single-particle and collective excitations
contribute equally to the SP integral (\ref{eq:21}) where the DF $%
\varepsilon (z,u)$\ is replaced by the effective one, $\varepsilon _{\mathrm{%
eff}}(z,u)$, given by Eq.~(\ref{eq:26}). Thus the LW equipartition rule
holds also in 2D treating the effective DF instead of $\varepsilon (z,u)$.
In this case it is straightforward to check that at $u\gg 1$\ the leading
order terms of the collective and single-particle excitations are given by $%
\Im _{p}\left( u\right) =\Im _{sp}\left( u\right) \simeq \pi /(4\alpha
u^{2}) $. The physical origin of the modification of the equipartition rule
in 2D is the change of the nature of the Coulomb potential (in Fourier space
it behaves as $\sim 1/k$ in 2D) and as a consequence the long-wavelength
dispersion relation: the plasma frequency behaves as $\sim k^{1/2}$ in this
limit. Technically this modification introduces an extra non-compensated $z$
variable as a prefactor in Eq.~(\ref{eq:3}), first line, which changes the
analytical properties of the DF. Introducing an effective DF (\ref{eq:26})
we formaly replace the 2D Coulomb potential by the 3D one without affecting
the polarizability of the 2D system. This recovers formally the 3D-type
dispersion relation with constant plasma frequency and hence the
equpartition rule.

\subsection{Low-velocity limit}
\label{sec:3.2}

Let us consider SP for slow projectiles, with $v\ll v_{F}$. A consequence of
the 3D linear response theory, confirmed by experiments, is that for ion
velocities $v$ low compared to the Fermi velocity $v_{F}$, the stopping
power is proportional to $v$ (see, e.g., the latest experiment \cite{moel02}%
). The coefficient of proportionality may be called a friction coefficient.
A similar linear behavior of the SP, $S\sim v$, is expected in 2D case Refs.~%
\cite{bre93,bre94,bre98,wan95,wang95}. Using analytical results obtained for
$\varepsilon _{\mathrm{RPA}}(z,u,\Gamma )$ the general expressions for SP
follow from Eqs.~(\ref{eq:1})-(\ref{eq:6}):
\begin{equation}
S\simeq \frac{2\Sigma _{0}Z^{2}}{\chi ^{2}}\lambda \int_{0}^{\infty }\frac{%
\Xi \left( z,\Gamma \right) z^{4}dz}{\left[ z^{2}+\chi ^{2}f\left( z\right) %
\right] ^{2}}=\frac{2\Sigma _{0}Z^{2}}{\chi ^{2}}\lambda \Re \left( \Gamma
,\chi ^{2}\right) ,
\label{eq:27}
\end{equation}%
where the dimensionless friction coefficient $\Re (\Gamma ,\chi ^{2})$
depends on the target properties and hence also on the dimensionless damping
parameter $\Gamma $. We have introduced the following functions
\begin{equation}
\Xi \left( z,\Gamma \right) =\frac{1}{\Gamma }\frac{f\left( z\right) \left[
2f\left( z\right) -\psi \left( z,\Gamma \right) \right] }{\psi \left(
z,\Gamma \right) } ,
\label{eq:28}
\end{equation}

\begin{equation}
\psi (z,\Gamma )=F_{1}(z,0,\Gamma )=2z+\frac{\Gamma }{z}\left[ \Phi
_{-}\left( z\right) -\Phi _{-}\left( -z\right) \right] -\left( z+1\right)
\Phi _{+}\left( z\right) -\left( z-1\right) \Phi _{+}\left( -z\right) ,
\label{eq:29}
\end{equation}

\begin{equation}
\Phi _{\pm }\left( z\right) =\frac{1}{\sqrt{2}}\sqrt{\sqrt{\frac{z^{2}\left(
z-1\right) ^{2}+\Gamma ^{2}}{z^{2}\left( z+1\right) ^{2}+\Gamma ^{2}}}\pm
\frac{z^{2}\left( z^{2}-1\right) +\Gamma ^{2}}{z^{2}\left( z+1\right)
^{2}+\Gamma ^{2}}} .
\label{eq:30}
\end{equation}%
The static screening function $f\left( z\right) $\ is determined from Eq.~(%
\ref{eq:11}).

\begin{figure}[tbp]
\includegraphics[width=14cm]{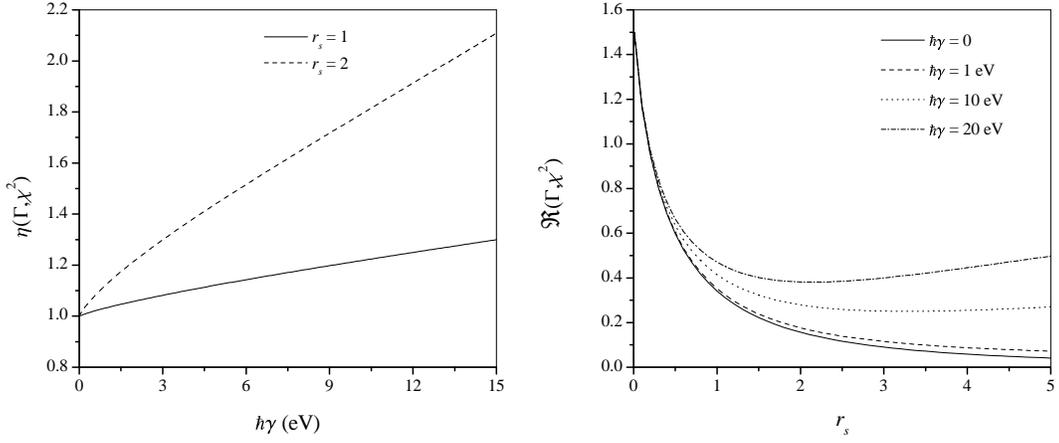}
\caption{Left panel, the ratio of the friction coefficients
with and without damping $\protect\eta (\Gamma ,\protect\chi ^{2})=\Re
(\Gamma ,\protect\chi ^{2})/\Re _{0}(\protect\chi ^{2})$ vs damping
parameter $\hbar \protect\gamma $ (in eV) for various densities. $r_{s}=1$
(solid line), $r_{s}=2$ (dashed line). Right panel, the friction coefficient
$\Re (\Gamma ,\protect\chi ^{2})$ vs density parameter $r_{s}$ for $\protect%
\gamma =0$ (solid line), $\hbar \protect\gamma =1$ eV (dashed line), $\hbar
\protect\gamma =10$ eV (dotted line) and $\hbar \protect\gamma =20$ eV
(dash-dotted line).}
\label{fig:4}
\end{figure}

When the damping vanishes ($\Gamma \rightarrow 0$) Eq.~(\ref{eq:29}) becomes
\begin{equation}
\psi \left( z,\Gamma \right) \rightarrow 2f\left( z\right) -\frac{2\Gamma }{%
\sqrt{1-z^{2}}}H\left( 1-z\right) +\mathrm{O}\left( \Gamma ^{2}\right) ,
\label{eq:31}
\end{equation}%
where $H\left( z\right) $ is the Heaviside unit-step function. Therefore
\begin{equation}
\left. \Xi \left( z,\Gamma \right) \right\vert _{\Gamma \rightarrow
0}\rightarrow \frac{1}{\sqrt{1-z^{2}}}H\left( 1-z\right)
\label{eq:32}
\end{equation}%
and from Eq.~(\ref{eq:27}) we find
\begin{eqnarray}
\left. \Re \left( \Gamma ,\chi ^{2}\right) \right\vert _{\Gamma \rightarrow
0} &=&\Re _{0}\left( \chi ^{2}\right) =\int_{0}^{1}\frac{z^{2}dz}{\left(
z+\chi ^{2}\right) ^{2}\sqrt{1-z^{2}}}  \label{eq:33} \\
&=&\frac{\pi }{2}+\frac{\chi ^{2}}{1-\chi ^{4}}\left[ 1+\frac{2(\chi ^{4}-2)%
}{\sqrt{|\chi ^{4}-1|}}\mathcal{G}\left( \chi ^{2}\right) \right]  \nonumber
\end{eqnarray}%
with
\begin{equation}
\mathcal{G}\left( x\right) =\left\{
\begin{array}{c}
\arctan \sqrt{\frac{x-1}{x+1}};\qquad x>1 \\
\frac{1}{2}\ln \left( \frac{1}{x}+\sqrt{\frac{1}{x^{2}}-1}\right) ;\qquad x<1%
\end{array}%
\right. .
\label{eq:34}
\end{equation}%
The last expressions (\ref{eq:33}) and (\ref{eq:34})\ are known results
derived previously within RPA in Refs.~\cite{bre93,wan95}. Interestingly, in
a low-velocity limit this SP completely agrees with the result obtained
within a binary collision approach Ref.~\cite{nag95}. In left panel of Fig.~%
\ref{fig:4} we show the ratio of the disorder-inclusive friction coefficient
$\Re (\Gamma ,\chi ^{2})$ and $\Re _{0}\left( \chi ^{2}\right) $\ vs damping
parameter $\hbar \gamma $ for two values of the density parameter $r_{s}=1$
and $r_{s}=2$. To gain more insight in right panel of Fig.~\ref{fig:4} we
show the friction coefficient $\Re (\Gamma ,\chi ^{2})$ vs $r_{s}$ for some
values of the damping parameter $\gamma $. As expected, the friction
coefficient and hence the SP at low velocities increase with an increasing
damping parameter $\gamma $; this was previously reported for 3D in Refs.~%
\cite{ner04,ner08,ash80}. The behavior of $\Re (\Gamma ,\chi ^{2})$ at fixed
$\gamma $ and at increasing density parameter $r_{s}$ is particularly
noteworthy. At small damping the friction coefficient decays monotonically
with $r_{s}$ while at large $\gamma $ it may also increase for large $r_{s}$%
. We will further discuss this behavior in Sec.~\ref{sec:4}.

The approximation (\ref{eq:27}) implies that the SP is proportional to
velocity. The velocity region in which the linear proportionality between SP
and the projectile velocity holds may be inferred from the numerical
calculations (see Sec.~\ref{sec:4}). It is seen from those results that the
approximation (\ref{eq:27}) remains quite accurate even when $\lambda $
becomes as large as $\sim 1$.

\subsection{High-velocity limit}
\label{sec:3.3}

Consider next the limit of large projectile velocities in the case of
strongly interacting DEG with the dielectric function Eqs.~(\ref{eq:14})-(%
\ref{eq:16}). In this limit the general expression (\ref{eq:1}) for
point-like projectiles with charge $Z$ moving in either interacting or free
electron gas reduces to the simple formula \cite{bre93}
\begin{equation}
S\simeq \frac{\pi \Sigma _{0}Z^{2}}{\chi ^{2}\lambda }=\frac{2\pi
^{2}n_{0}Z^{2}e^{4}}{\hbar v}
\label{eq:35}
\end{equation}%
which does not contain the gas electron mass $m$ anymore; $m$ and also the
effects of electron-electron interactions appear only in the higher terms of
the expansion. The other main discrepancy between the 2D and the 3D results
is that the stopping power decreases as $1/v$ instead of behaving as $\ln
(v)/v^{2}$ in the 3D case. In the presence of interactions the next order
terms are shown to be significantly modified. We derive below a generalized
expression for SP, in a high-velocity limit, for point-like ions. In order
to show how SP in a high-velocity limit is affected we consider expression (%
\ref{eq:1}) rewritten as follows:
\begin{equation}
S=\frac{2\Sigma _{0}Z^{2}}{\chi ^{2}\lambda }\int_{0}^{\lambda }\frac{%
\Lambda \left( u\right) du}{\sqrt{\lambda ^{2}-u^{2}}} ,
\label{eq:36}
\end{equation}%
where
\begin{equation}
\Lambda \left( u\right) =\frac{4\alpha }{\pi }u\Im \left( u\right) =\frac{%
4\alpha }{\pi }u\int_{0}^{\infty }\mathrm{Im}\frac{-1}{\varepsilon \left(
z,u\right) }zdz
\label{eq:37}
\end{equation}%
and $\Im \left( u\right) $ is the total contribution of the collective and
single-particle excitations to the SP integral defined in Sec.~\ref{sec:3.1}
(see Eqs.~(\ref{eq:21})-(\ref{eq:23})). For further progress it is
imperative to calculate the asymptotic behavior of the function $\Lambda
\left( u\right) $ at $u\rightarrow \infty $. For collective and
single-particle excitations these asymptotic forms are given by Eqs.~(\ref%
{eq:24}) and (\ref{eq:25}), respectively. Using these expressions we arrive
at
\begin{equation}
\Lambda \left( u\right) =1+\frac{C_{2}}{u^{2}}+\frac{C_{3}}{u^{3}}+\mathrm{O}%
\left( u^{-4}\right)
\label{eq:38}
\end{equation}%
for $u\rightarrow \infty $ and with the expansion coefficients
\begin{equation}
C_{2}=\frac{1}{4},\qquad C_{3}=\frac{3}{2\alpha }\left[ 1-\varkappa
_{1}\left( r_{s}\right) \right] .
\label{eq:coef}
\end{equation}%
Here the parameter $\varkappa _{1}\left( r_{s}\right) $\ depends on the
exchange-correlation interactions and is given explicitly in Appendix~\ref%
{sec:app2}.

Below we calculate the SP up to the order $\mathrm{O}(v^{-4})$\ thus
neglecting the terms with $\mathrm{O}\left( \lambda ^{-5}\right) $. First
the SP (\ref{eq:36}) can be represented in the equivalent form
\begin{eqnarray}
S &=&\frac{\pi \Sigma _{0}Z^{2}}{\chi ^{2}\lambda }\left\{ 1+\frac{h_{1}}{%
\lambda }+\frac{1}{2\pi \lambda ^{2}}\left[ 1-\frac{1}{\lambda +\sqrt{%
\lambda ^{2}-1}}-\Phi _{2}\left( \lambda \right) \right] \right.
\label{eq:39} \\
&&\left. +\frac{C_{3}}{\pi \lambda ^{3}}\left[ \frac{3}{2}+\frac{1}{4C_{3}}-%
\frac{\lambda }{\lambda +\sqrt{\lambda ^{2}-1}}+\ln \left( \lambda +\sqrt{%
\lambda ^{2}-1}\right) +\Phi _{1}\left( \lambda \right) \right] \right\}
\nonumber
\end{eqnarray}%
which is convenient for further calculations. Here $h_{1}$ is a constant
\begin{equation}
h_{1}=\frac{2}{\pi }\int_{0}^{\infty }\left[ \Lambda \left( u\right) -1%
\right] du
\label{eq:h1}
\end{equation}%
and the other quantities are function of the ion velocity:
\begin{eqnarray}
\Phi _{1}\left( \lambda \right) &=&\frac{2\lambda ^{2}}{C_{3}}%
\int_{0}^{1}\left( \frac{1}{\sqrt{1-u^{2}/\lambda ^{2}}}-1\right) \left[
\Lambda \left( u\right) -1\right] du  \label{eq:40} \\
&&+\frac{2\lambda ^{2}}{C_{3}}\int_{1}^{\lambda }\left( \frac{1}{\sqrt{%
1-u^{2}/\lambda ^{2}}}-1\right) \left[ \Lambda \left( u\right) -1-\frac{C_{2}%
}{u^{2}}-\frac{C_{3}}{u^{3}}\right] du-\frac{1}{2}-\frac{1}{4C_{3}},  \nonumber
\end{eqnarray}

\begin{equation}
\Phi _{2}\left( \lambda \right) =4\lambda \int_{\lambda }^{\infty }\left[
\Lambda \left( u\right) -1\right] du .
\label{eq:41}
\end{equation}%
For the derivation of Eq.~(\ref{eq:39}) we have used some elementary
integrals \cite{gra80}. In Appendix~\ref{sec:app2} we prove that $h_{1}=0$,
see Eq.~(\ref{eq:app9}). This relation can be regarded as another SP sum
rule for an interacting DEG.

For a calculation of the SP up to fourth order $v^{-4}$ we need the
asymptotic behavior of $\Phi _{2}\left( \lambda \right) $\ up to the first
order ($v^{-1}$) which can be obtained from Eqs.~(\ref{eq:38}) and (\ref%
{eq:41})
\begin{equation}
\Phi _{2}\left( \lambda \right) =1+\frac{2C_{3}}{\lambda }+\mathrm{O}\left(
\lambda ^{-2}\right) ,
\label{eq:42}
\end{equation}%
and only the leading term of $\Phi _{1}\left( \lambda \right) $. We denote
this leading term by $\left. \Phi _{1}\left( \lambda \right) \right\vert
_{\lambda \rightarrow \infty }=\ln h_{2}$ and using Eq.~(\ref{eq:40}) we
obtain
\begin{equation}
\ln h_{2}=\frac{1}{C_{3}}\left\{ \int_{0}^{1}\left[ \Lambda \left( u\right)
-1\right] u^{2}du+\int_{1}^{\infty }\left[ \Lambda \left( u\right) -1-\frac{%
C_{2}}{u^{2}}-\frac{C_{3}}{u^{3}}\right] u^{2}du-\frac{1}{4}\right\} -\frac{1%
}{2} .
\label{eq:h2}
\end{equation}%
The coefficient $\ln h_{2}$ is explicitly evaluated in Appendix~\ref%
{sec:app2} and entirely depends on the density parameter $r_{s}$, see Eq.~(%
\ref{eq:app15}). Thus substituting Eqs.~(\ref{eq:42}) and (\ref{eq:h2}) into
(\ref{eq:39}) and setting $h_{1}=0$ we finally obtain
\begin{equation}
S\simeq \frac{\pi \Sigma _{0}Z^{2}}{\chi ^{2}\lambda }\left[ 1+\frac{C_{3}}{%
\pi \lambda ^{3}}\ln \left( 2h_{2}\lambda \right) \right] .
\label{eq:43}
\end{equation}%
It is seen that in the correction term (the second term in Eq.~(\ref{eq:43}%
)) the mass of electron enters through the Fermi velocity $v_{F}=\hbar
k_{F}/m $. A limit to the non-interacting DEG is performed by taking the
limit $\varkappa _{1}\left( r_{s}\right) \rightarrow 0$, i.e. setting $%
C_{3}=3/2\alpha =(3\sqrt{2}/4)r_{s}$\ (see Eq.~(\ref{eq:coef})). In this
limit the coefficient $\ln h_{2}$ is given by Eq.~(\ref{eq:app16}). In the
general case of non-vanishing exchange-correlation interactions it is too
difficult to draw some conclusions from Eq.~(\ref{eq:43}) about how these
interactions affect the high-velocity SP. Numerical calculations of Refs.~%
\cite{wan95,wang95} show that these interactions strongly increase the SP up
to the intermediate velocity range with $v\sim v_{F}$. We support this
conclusion by our own calculations (not shown here) which also indicate that
the asymptotic SP (\ref{eq:43}) remains quite accurate also in the
intermediate velocity range.

\begin{figure}[tbp]
\includegraphics[width=14cm]{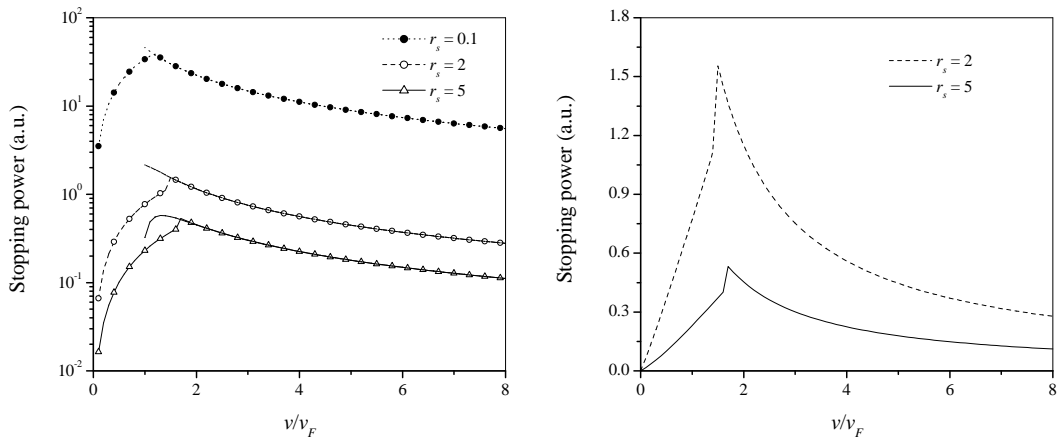}
\caption{Left panel, the SP (in a.u. and in logarithmical
scale) of a proton vs $v/v_{F}$ moving in an interacting electron gas for $%
r_{s}=0.1$ (dotted line), $r_{s}=2$ (dashed line), $r_{s}=5$ (solid line).
The lines with and without symbols correspond to the numerical evaluation of
Eq.~(\protect\ref{eq:1}) with Eqs.~(\protect\ref{eq:14})-(\protect\ref{eq:16}%
) and asymptotic expression (\protect\ref{eq:43}), respectively. Right
panel, the dashed and solid lines (with symbols) from the left panel in
linear scale.}
\label{fig:5}
\end{figure}

We close this section with the following two remarks. First, the high-velocity
SP Eq.~(\ref{eq:43}) is also valid for a general LFC function $G(k)$. The derivations
above and in Appendix~\ref{sec:app2} show that only the asymptotic values of
$G(k)$ at $k\to \infty$ and $k\to 0$ contribute to Eq.~(\ref{eq:43}). At
short wavelengths $G(k\to \infty )=G_{\infty }(r_{s})$ is constant
(see, e.g., Ref.~\cite{gol93}). At long wavelengths the LFC function behaves
as $G(k\to 0)\simeq \kappa (r_{s})k/k_{F}$, where the constant $\kappa (r_{s})$
is related to the compressibility of a 2D electron gas through compressibility
sum rule. The latter for a 3D electron gas is discussed in \cite{sin81}, and
for a 2D electron gas in \cite{ham03}. Thus in the general case of arbitrary $G(k)$ the quantities
$\varkappa _{1}(r_{s})$ and $\varkappa _{2}(r_{s})$ in Eq.~(\ref{eq:43}) are
replaced by $\varkappa _{1}(r_{s})=(1/3)G_{\infty }(r_{s})$ and $\varkappa
_{2}(r_{s})=G_{\infty }(r_{s})/4\kappa (r_{s})$, respectively. Second, a
similar procedure is applicable to evaluate the high-velocity
corrections also for a disordered 2D electron gas. While the high-velocity
SP (\ref{eq:43}) does not contain the terms of the second $v^{-2}$ and
third $v^{-3}$ orders, some preliminary investigations by us show that for
a disordered DEG this SP involves also the terms of the order $B_{1}v^{-2}$,
$B_{2}v^{-2}\ln v$ and $B_{3}v^{-2}\ln ^{2}v$, where the constants $B_{1}$,
$B_{2}$ and $B_{3}$ depend on $\gamma $. Therefore the corrections to the
high-velocity SP would be much more sensitive to the ion velocity than those
predicted by Eq.~(\ref{eq:43}).

\section{Numerical Calculations}
\label{sec:4}

\begin{figure}[tbp]
\includegraphics[width=8cm]{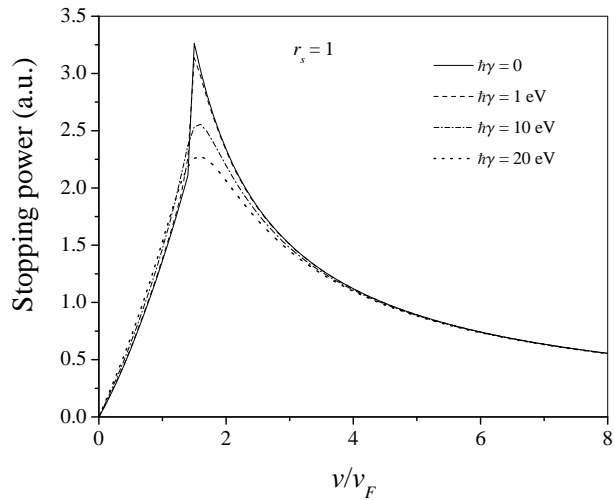}
\caption{The SP (in a.u.) of a proton vs $v/v_{F}$ moving in
a disordered electron gas for $r_{s}=1$, $\protect\gamma =0$ (solid line), $%
\hbar \protect\gamma =1$~eV (dashed line), $\hbar\protect\gamma =10$~eV
(dash-dotted line), $\hbar\protect\gamma =20$~eV (dotted line).}
\label{fig:6}
\end{figure}

\begin{figure}[tbp]
\includegraphics[width=8cm]{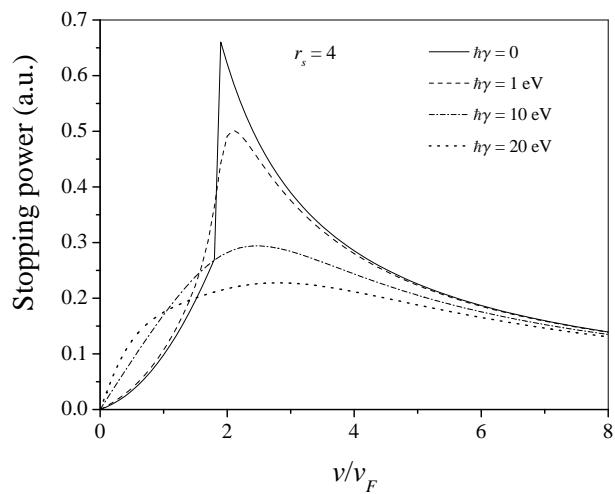}
\caption{Same as in Fig.~\protect\ref{fig:6} but for $r_{s}=4$.}
\label{fig:7}
\end{figure}

Using the theoretical results obtained in Secs.~\ref{sec:2} and \ref{sec:3},
we present here the results of our numerical calculations of stopping power
for a 2D target material with the wide range of the density parameter, $%
0.1\leqslant r_{s}\leqslant 5$. The parameter $r_{s}$\ varies from the small
(free DEG) up to the large (strongly interacting DEG) values. As examples of
2D target material we have considered two models. An interacting DEG whose
linear response function includes the exchange-correlation effects via LFC
and is given by Eqs.~(\ref{eq:14})-(\ref{eq:16}). This case has been
investigated previously in Refs.~\cite{wan95,wang95}. In Fig.~\ref{fig:5}
left panel we compare the exact (the lines with symbols) and asymptotic (the
lines without symbols) SPs calculated from Eqs.~(\ref{eq:1}), (\ref{eq:14})-(%
\ref{eq:16}) and (\ref{eq:43}), respectively. It is seen that the asymptotic
expression (\ref{eq:43})\ is very accurate and at $v\gtrsim v_{F}$
practically coincides with exact SP. In general we have found that the
higher order correction in Eq.~(\ref{eq:43}) (the second term) is small
compared to the leading term. However, the role of this term becomes more
and more pronounced with increasing the density parameter $r_{s}$, i.e. with
increasing the exchange-correlation interactions. We have also compared our
numerical calculations with the results obtained by Wang and Ma~\cite%
{wan95,wang95}. Two major differences have been found. First, the LFC
dielectric function (\ref{eq:14}) for a fully degenerate electron gas
predicts a threshold ion velocity for plasmon excitations. In view of the
discussion in Sec.~\ref{sec:2.2} the plasmons are excited at $\lambda
>\lambda _{c}$, where the critical (dimensionless) velocity $\lambda _{c}$
is the minimum value of the dispersion function $u_{r}(z)$, Eq.~(\ref{eq:17}%
), and can be found from the equations $\lambda _{c}=u_{r}(z_{\min })$ with $%
u_{r}^{\prime }(z_{\min })=0$ (see also Table~\ref{tab:1}). The velocity
threshold changes sufficiently the slope of the SP and at $\lambda =\lambda
_{c}$ one expects a characteristic discontinuity of the derivative of the SP
(the SP itself remains naturally continous at $\lambda =\lambda _{c}$). In
contrast to Refs.~\cite{wan95,wang95} this feature is clearly visible in
Fig.~\ref{fig:5}, right panel (see also the solid lines in Figs.~\ref{fig:6}
and \ref{fig:7}). Such behavior of the SP at $\lambda =\lambda _{c}$ has
been observed previously in 3D (see, e.g., \cite{ner00} and references
therein).

Second, we have found that for the same conditions (i.e. for the same $r_{s}$%
) the SP in our case is considerably smaller near maximum than those
obtained in Ref.~\cite{wan95}. Moreover, there is no agreement between the
results obtained in Refs.~\cite{wan95} and \cite{wang95}, e.g. for $r_{s}=1$
and $r_{s}=5$, where Ref.~\cite{wan95} predicts in whole velocity range much
larger SP than the latter. Apparently this is because the polarizability of
the free-electron gas employed in Ref.~\cite{wan95} somewhat differs from
original expression derived by Stern \cite{ste67} (see also Eq.~(\ref{eq:15}%
) with Eqs.~(\ref{eq:7}), (\ref{eq:8})); the algebraic square roots in Eqs.~(%
\ref{eq:7}) and (\ref{eq:8}) are missing in Ref.~\cite{wan95}. These square
roots are recovered in Ref.~\cite{wang95} but nevertheless one of two
plasmon branches is ignored as discussed in Sec.~\ref{sec:2.2} which may
yield smaller value of the SP.

Within the second model the target material is modelled as an electron gas
whose linear response function, within RTA, is given by Eqs.~(\ref{eq:2})-(%
\ref{eq:6}) with $\gamma $ as a model damping parameter. In Figs.~\ref{fig:6}
and \ref{fig:7} based on numerical calculations we choose four values of $%
\gamma $: $\hbar \gamma =0$ (solid lines), $\hbar \gamma =1$ eV (dashed
lines), $\hbar \gamma =10$ eV (dash-dotted lines), and $\hbar \gamma =20$ eV
(dotted lines). The values $0<\hbar \gamma <1$ eV are comparable with the
damping parameters (related inversely to the collision times) in some 3D
metal targets, e.g., Al for which $\hbar \gamma $ can be $\sim 0.1$ eV. The
last values $\hbar \gamma =10$ and $20$ eV correspond to the damping
parameter, e.g. in carbon. It is seen from Figs.~\ref{fig:6} and \ref{fig:7}
that the SP is broadened with increasing damping (i.e. with increasing $%
\gamma $) and this effects is more pronounced for small densities (i.e. at
large $r_{s}$). Of course, the value $r_{s}=4$\ in Fig.~\ref{fig:7} is
somewhat far beyond the RPA employed for deriving the dielectric function
Eqs.~(\ref{eq:2})-(\ref{eq:6}). However, treating this case as a qualitative
example we look for some complementary information about the effect of
disorder at large $r_{s}$. In particular, at either vanishing or small
damping with $\gamma =0$ and $\hbar \gamma =1$~eV, respectively, Fig.~\ref%
{fig:7} predicts a modification of the linear friction law (see Eq.~(\ref%
{eq:27})) which now approximately behaves as $\sim v^{3}$. This is $v^{3}$
law obtained e.g. in Ref.~\cite{pet91} within linear response theory for
classical 3D plasma and supported by the numerical simulations \cite{zwi99}.
According to Fig.~\ref{fig:4} (right panel) at small $\gamma $\ the linear
friction coefficient decays with $r_{s}$ and may be smaller than the cubic
friction coefficient ($\sim v^{3}$). However at strong damping the linear
friction coefficient again becomes dominant and the SP at small velocities
behaves as $\sim v$ (cf. Figs.~\ref{fig:4} (right panel) and \ref{fig:7}).

\section{Summary and concluding remarks}
\label{sec:sum}

In this paper we have presented a theoretical study of the stopping power of
point ion projectile in a degenerate 2D electron gas. The later is modelled
within two different approaches namely (i) as a system containing disorder
due to, e.g., electron-impurity interactions and (ii) including
exchange-correlation interactions of the electrons. In the course of this
study we have also derived some analytical results for the
disorder-inclusive RPA linear response function and for the corresponding
plasmon dispersion relations. These analytical results go beyond those
obtained previously in Refs.~\cite{giu84,gel86,gor91} within small-$k,\omega
$ approximation. Also for the model (ii) we have found an exact dispersion
relations. After a general introduction to the SP of an ion in Sec.~\ref%
{sec:2} theoretical calculations of SP based on the linear response theory
and using the models (i) and (ii) are discussed in Sec.~\ref{sec:3}. A
number of limiting and asymptotic regimes of low- and high-velocities and
vanishing damping have been studied. These approximate expressions are well
supported by our numerical calculations. Special attention has been paid to
the equipartition sum rule in 2D. In Sec.~\ref{sec:3} employing the model
(ii), i.e. LFC dielectric function for an interacting DEG, we have shown
that this rule does not necessarily hold in 2D and may be satisfied
introducing an effective dielectric function~(\ref{eq:26}). The theoretical
expressions for a number of physical quantities derived in this paper lead
to a detailed presentation, in Secs.~\ref{sec:2}-\ref{sec:4}, of a
collection of data through figures on SP, friction coefficient and the
dispersion relations. For the damping parameter, we have chosen a wide range
of values $0\leqslant \hbar \gamma \leqslant 20$ eV; the damping parameters
for some 3D metal and semiconductor targets fall within this range. The
results we have presented demonstrate that with regard to several physical
quantities of primary interest the difference between RTA and usual RPA
without damping is significant.

It is of particular interest to study the high-velocity limit for the SP of
an ion beam. Such asymptotic expressions contain some useful information on
a projectile ion structure factor and specially on the target medium
properties. Eq.~(\ref{eq:43}) with Eq.~(\ref{eq:app15}) which are a
generalization of the asymptotic formula obtained in Refs.~\cite%
{bre93,bre94,bre98,wan95} can be used for analyses of experimental data on
high-energy beam interactions with 2D target material. We note that the
analytical method developed here for the derivation of high-velocity SP is
general and may be applied within a linear response treatment for other
types of projectiles, e.g. extended multicharged ions, as well as for any
particular form of the linear response function $\varepsilon \left(
z,u\right) $ for the target material. For given target material this
approach requires only the asymptotic form of the plasmon dispersion
relation at high $u=\omega /kv_{F}$ and the frequency moments of the energy
loss function. For a disordered DEG (model (i)), however, some modifications
occur when one includes the damping in the DF. For instance, at large
frequencies the energy loss function $\mathrm{Im}[-1/\varepsilon (z,u)]$ for
a disordered DEG behaves as $\gamma \omega _{p}^{2}(k)/\omega ^{3}$ and
obviously the third frequency moment of this function does not exist, see
Eq.~(\ref{eq:app10}). This requires some additional investigation of the
third moment sum rule for this case which in turn is important for
evaluation of high-velocity SP, see Sec.~\ref{sec:3.3}.

We shall make some brief remarks on the RTA in the linear response function.
In the present study the disorder-inclusive linear response function
containing in the RTA has been considered only in RPA. Going beyond RPA with
electron-electron interaction and disorder treated at the same microscopic
level is a difficult task. We may mention that recently the linear response
function in 3D has been considered in RTA which conserves the particle
number, momentum and energy \cite{mor00,atw02} (see also references
therein). We intend to extend this model with fully conserving (number,
momentum and energy) linear response function for 2D electron gas.

In our calculations of SP and related quantities we have modelled the
disordered 2D target medium as an electron gas whose linear response
function is constructed in RTA in order to include scattering of electrons
with disorder impurities. The numerical values of the phenomenological
quantity $\gamma $ used in our calculations are within a physically expected
range for the specific target medium. In principle $\gamma $ can be
calculated to varying degrees of approximations. In the simplest
approximation, its inverse can be calculated through Fermi's golden rule for
a model electron-impurity potential. This may allow us to see how SP and
related quantities depend on the target properties through their influence
on $\gamma $.

We expect our theoretical findings to be useful in experimental
investigations of ion beam energy losses in solids. One of the improvements
of our model will be to include some short-range correlation in the linear
response function. Another interesting issue not considered here in detaile
is the effective DF~(\ref{eq:26}) for 2D interacting DEG. Our goal is to
find physical motivation and basis for this type of DF. A study of this and
other aspects will be reported elsewhere.

\begin{acknowledgments}
The work of H.B.N. has been partially supported by the Armenian Ministry of
Higher Education and Science Grant No.~87.
\end{acknowledgments}

\appendix

\section{Dielectric function of disordered electron gas}
\label{sec:app1}

In this Appendix we give an alternative derivation of the disorder-inclusive
DF which is valid in the entire complex $\omega ,k$-plane. Since we are
going to compare our results with previous derivations in Refs.~\cite%
{giu84,gel86,gor91} here we use the usual energy ($\omega $) and momentum ($%
k $) variables. Performing the $q$ and $\theta $ integrations in Eq.~(\ref%
{eq:3}) without splitting this expression into real and imaginary parts, for
arbitrary $\omega $ and $k$ complex variables we obtain
\begin{equation}
\varepsilon _{\mathrm{RPA}}\left( k,\omega \right) =1+\frac{2}{ka_{0}}\left[
1-\frac{2\omega _{+}}{\sqrt{\left( \omega _{+}-\omega _{k}\right)
^{2}-k^{2}v_{F}^{2}}+\sqrt{\left( \omega _{+}+\omega _{k}\right)
^{2}-k^{2}v_{F}^{2}}}\right] ,
\label{eq:ap1}
\end{equation}%
where $\omega _{+}=\omega +i\gamma $, $\omega _{k}=\hbar k^{2}/2m$. Here $%
\hbar \omega _{k}$\ is the single-particle energy and $\lambda _{\mathrm{TF}%
}=a_{0}/2$ plays a role of the Thomas-Fermi screening length which is
constant in 2D case \cite{fet73}. The multi-valued functions in Eq.~(\ref%
{eq:ap1}) must be understood in the following way. (i) The imaginary parts
of the square roots are positive. (ii) The signs of the real parts of the
square roots with $\omega _{+}\pm \omega _{k}$ are taken with the sign of
the expressions
\begin{equation}
\frac{\left\vert u_{\pm }\right\vert }{u_{\pm }}=\frac{\left\vert \omega \pm
\hbar k^{2}/2m\right\vert }{\omega \pm \hbar k^{2}/2m} .
\label{eq:ap2}
\end{equation}%
These two conditions completely fix uniquely the values of the square roots
entered in Eq.~(\ref{eq:ap1}). The full number-conserving DF is now
evaluated using Mermin-Das formula, Eq.~(\ref{eq:2})
\begin{equation}
\varepsilon \left( k,\omega \right) =1+\frac{2}{ka_{0}}\left\{ 1-\frac{%
\omega +i\gamma Q\left( k,\omega \right) }{P\left( k,\omega \right) +i\gamma %
\left[ Q\left( k,\omega \right) -1\right] }\right\} .
\label{eq:ap3}
\end{equation}%
Here
\begin{equation}
P\left( k,\omega \right) =\frac{1}{2}\left[ \sqrt{\left( \omega _{+}-\omega
_{k}\right) ^{2}-k^{2}v_{F}^{2}}+\sqrt{\left( \omega _{+}+\omega _{k}\right)
^{2}-k^{2}v_{F}^{2}}\right] ,
\label{eq:ap4}
\end{equation}

\begin{equation}
Q\left( k,\omega \right) =\frac{2}{ka_{0}}\frac{1}{\omega _{+}}\left[
P\left( k,\omega \right) -\omega _{+}\right] \left[ \nu \left( k\right) -1-%
\frac{ka_{0}}{2}\right] ,
\label{eq:ap5}
\end{equation}

\begin{equation}
\nu \left( k\right) =\frac{\varepsilon _{\mathrm{RPA}}(k,0)}{\varepsilon _{%
\mathrm{RPA}}(k,0)-1}=1+\frac{ka_{0}}{2}\frac{k/2k_{F}}{f\left(
k/2k_{F}\right) }
\label{eq:ap6}
\end{equation}%
and the function $f\left( z\right) $ has been introduced by Eq.~(\ref{eq:11}%
).

Now let us consider the limit of small momentum-energy transfers, i.e. we
assume that $k\ll 2k_{F}$ and $\hbar \omega \ll E_{F}$. In this case $f(z)=z$
and the function $Q$ in Eq.~(\ref{eq:ap5}) vanishes. In addition neglecting
the single-particle energies $\hbar \omega _{k}$ in Eqs.~(\ref{eq:ap1}), (%
\ref{eq:ap3}) and (\ref{eq:ap4}) we obtain
\begin{equation}
\varepsilon _{\mathrm{RPA}}\left( k,\omega \right) \simeq 1+\frac{2}{ka_{0}}%
\left( 1-\frac{\omega _{+}}{\sqrt{\omega _{+}^{2}-k^{2}v_{F}^{2}}}\right)
\label{eq:ap7}
\end{equation}%
and
\begin{equation}
\varepsilon \left( k,\omega \right) \simeq 1+\frac{2}{ka_{0}}\left[ 1-\frac{%
\omega }{\sqrt{\omega _{+}^{2}-k^{2}v_{F}^{2}}-i\gamma }\right] .
\label{eq:ap8}
\end{equation}%
These are precisely the same DFs obtained previously in Refs.~\cite%
{giu84,gel86,gor91} which used the same small $k,\omega$-approximation
limits of the more general expressions (\ref{eq:ap1}) and (\ref{eq:ap3}).
Note that the DFs (\ref{eq:ap7}) and (\ref{eq:ap8}) are the quasiclassical
limits of the more general expressions (\ref{eq:ap1}) and (\ref{eq:ap3}),
respectively. Therefore they can be alternatively derived from a classical
kinetic equation within RTA with the Fermi distribution function as an
unperturbed state.

\section{Evaluation of the parameters $h_{1}$ and $h_{2}$}
\label{sec:app2}

In this Appendix we give detail derivation of the parameters $h_{1}$\ and $%
h_{2}$\ which contributes to the high-velocity SP of an interacting 2D
electron gas, Eqs.~(\ref{eq:h1}) and (\ref{eq:h2}), respectively. First we
write Eq.~(\ref{eq:h1}) in another but equivalent form, $h_{1}=\left. \wp
\left( s\right) \right\vert _{s\rightarrow \infty }$, where
\begin{eqnarray}
\wp \left( s\right) &=&\frac{2}{\pi }\left[ \int_{0}^{s}\Lambda \left(
u\right) du-s\right]  \label{eq:app4} \\
&=&\frac{8\alpha }{\pi ^{2}}\left[ \int_{s}^{\infty }zdz\int_{0}^{s}L\left(
z,u\right) udu-\int_{0}^{s}zdz\int_{s}^{\infty }L\left( z,u\right) udu\right]
.  \nonumber
\end{eqnarray}%
Here $L(z,u)=\mathrm{Im}[-1/\varepsilon (z,u)]$\ is the energy loss
function. For derivation of Eq.~(\ref{eq:app4}) the Bethe sum rule (the
first frequency moment of the energy loss function) in variables $z$ and $u$
has been used (see, e.g., Refs.~\cite{bre93,bre94,bal06})
\begin{equation}
\int_{0}^{\infty }\mathrm{Im}\frac{-1}{\varepsilon (z,u)}udu=\frac{\pi \chi
^{2}}{4z}=\frac{\pi }{4\alpha z} .
\label{eq:app5}
\end{equation}%
Assuming that the upper cutoff $s$ is large enough, $s\gg 1$, Eq.~(\ref%
{eq:app4}) can be written in explicit form
\begin{eqnarray}
\wp \left( s\right) &=&\frac{8\alpha }{\pi ^{2}}\left[ \int_{s}^{s+1}zdz%
\int_{z-1}^{s}L\left( z,u\right) udu-\int_{s-1}^{s}zdz\int_{s}^{z+1}L\left(
z,u\right) udu\right]  \label{eq:app6} \\
&&-\frac{16\alpha ^{2}}{\pi }\int_{0}^{z_{r}\left( s\right) }\frac{g\left(
z\right) }{\left\vert g\left( z\right) \right\vert }\frac{z^{3}g^{2}\left(
z\right) u_{r}\left( z\right) dz}{\left\vert \phi _{r}\left( z\right)
\right\vert }.  \nonumber
\end{eqnarray}%
Here $u_{r}(z)$ is the solution of the dispersion equation for an
interacting DEG, Eq.~(\ref{eq:17}), and we have introduced a lower cutoff
parameter $z_{r}\left( s\right) $\ which $z_{r}\left( s\right) \rightarrow 0$
at $s\rightarrow \infty $. Also we have introduced the function $\phi
_{r}\left( z\right) $ which is given by
\begin{equation}
\phi _{r}\left( z\right) =\left. \frac{\partial }{\partial u}f_{1}\left(
z,u\right) \right\vert _{u=u_{r}\left( z\right) }=\frac{u_{r}\left( z\right)
-z}{\sqrt{\left[ u_{r}\left( z\right) -z\right] ^{2}-1}}-\frac{u_{r}\left(
z\right) +z}{\sqrt{\left[ u_{r}\left( z\right) +z\right] ^{2}-1}}.
\label{eq:app7}
\end{equation}%
The last term in Eq.~(\ref{eq:app6}) is the contribution of the collective
excitations and hence the function $f_{1}(z,u)$\ in Eq.~(\ref{eq:app7}) is
defined in the domain $0<z<u-1$.(or $u>z+1$). Without lose of the generality
we chose as a lower cutoff (i.e. $z_{r}\left( s\right) $) a function which
is inverse to $u_{r}\left( z\right) $. Using Eq.~(\ref{eq:17}) it is
straightforward to calculate the asymptotic behavior of this function at
large $u$. It behaves as
\begin{equation}
z_{r}\left( u\right) =\frac{1}{2\alpha u^{2}}\left( 1+\frac{A_{1}}{u^{2}}+%
\frac{A_{2}}{u^{4}}+\frac{A_{3}}{u^{6}}+...\right) ,
\label{eq:app1}
\end{equation}%
where the expansion coefficients are given by
\begin{eqnarray}
A_{1} &=&\frac{3}{4}-\frac{1}{2}\varkappa \left( r_{s}\right) ,  \nonumber \\
A_{2} &=&\frac{5}{8}\left[ 1-\varkappa \left( r_{s}\right) \right] \left[ 1-%
\frac{1}{2}\varkappa \left( r_{s}\right) \right] +\frac{\varkappa \left(
r_{s}\right) }{16}\left[ 3-\varkappa \left( r_{s}\right) \right] ,
\label{eq:app2} \\
A_{3} &=&\frac{35}{64}+\frac{1}{4\alpha ^{2}}+\frac{\varkappa \left(
r_{s}\right) }{16}\left[ \varkappa _{0}^{2}\left( r_{s}\right) -2\varkappa
^{2}\left( r_{s}\right) +9\varkappa \left( r_{s}\right) -\frac{29}{2}\right],
\nonumber
\end{eqnarray}%
with $\varkappa \left( r_{s}\right) =G_{0}\left( r_{s}\right) /[\alpha
G_{12}\left( r_{s}\right) ]$ and $\varkappa _{0}\left( r_{s}\right)
=G_{22}\left( r_{s}\right) /[\alpha G_{12}\left( r_{s}\right) ]$.

Since at small $z$ the functions $u_{r}\left( z\right) $ and $\phi
_{r}\left( z\right) $\ behave as $u_{r}\left( z\right) \simeq (2\alpha
z)^{-1/2}$ and
\begin{equation}
\phi _{r}\left( z\right) \simeq 4\alpha \sqrt{2\alpha }z^{5/2}\left\{ 1+%
\frac{3\alpha z}{4}\left[ 1+2\varkappa \left( r_{s}\right) \right] +\mathrm{O%
}\left( z^{2}\right) \right\} ,
\label{eq:app8}
\end{equation}%
respectively, at $s\rightarrow \infty $ the plasmon contribution in Eq.~(\ref%
{eq:app6}) vanishes as $\sim z_{r}\left( s\right) \sim s^{-2}\rightarrow 0$.
For calculation of the first two terms in Eq.~(\ref{eq:app6})
(single-particle contributions) we first make a substitution of the
integration variables, $z\rightarrow z+s$ and $u\rightarrow u+s$. At $%
s\rightarrow \infty $ the remaining expression behaves as $\wp \left(
s\right) \simeq -1/(2\pi s)\rightarrow 0$. Thus at $s\rightarrow \infty $, $%
\wp \left( s\right) \rightarrow 0$ and
\begin{equation}
h_{1}=\left. \wp \left( s\right) \right\vert _{s\rightarrow \infty }=0.
\label{eq:app9}
\end{equation}

For calculation of the coefficient $h_{2}$\ it is imperative to evaluate the
third moment of the energy loss function. In 2D and in general case this has
been done in Ref.~\cite{bal06}. In the present context of an interacting
electron gas with DF (\ref{eq:14}) and (\ref{eq:15}) this moment is given by
\begin{equation}
\int_{0}^{\infty }\mathrm{Im}\frac{-1}{\varepsilon \left( z,u\right) }%
u^{3}du=\frac{\pi }{8\alpha }\left\{ \frac{1}{\alpha z^{2}}\left[ 1-G\left(
z\right) \right] +\frac{3}{2z}+2z\right\} .
\label{eq:app10}
\end{equation}%
As we have done above we introduce now a new function through the relation $%
h_{2}=\left. \Im \left( s\right) \right\vert _{s\rightarrow \infty }$, where
\begin{equation}
\ln \Im \left( s\right) =\frac{1}{C_{3}}\int_{0}^{s}\Lambda \left( u\right)
u^{2}du-\ln s-\frac{s}{4C_{3}}-\frac{s^{3}}{3C_{3}}-\frac{1}{2}
\label{eq:app11}
\end{equation}%
and $C_{3}$ is given by Eq.~(\ref{eq:coef}). Employing the relation (\ref%
{eq:app10}) for the third frequency moment we obtain
\begin{eqnarray}
\ln \Im \left( s\right) &=&\frac{1}{3\left[ 1-\varkappa _{1}\left(
r_{s}\right) \right] }\left[ \ln \left( 2\alpha \right) +3\varkappa
_{1}\left( r_{s}\right) \ln \varkappa _{2}\left( r_{s}\right) \right] -\frac{%
1}{2}  \label{eq:app12} \\
&&-\frac{1}{2\left[ 1-\varkappa _{1}\left( r_{s}\right) \right] }U\left(
s\right) ,  \nonumber
\end{eqnarray}%
where $\varkappa _{1}\left( r_{s}\right) =G_{0}\left( r_{s}\right)
/[3G_{22}\left( r_{s}\right) ]$, $\varkappa _{2}\left( r_{s}\right)
=G_{12}\left( r_{s}\right) /[2G_{22}\left( r_{s}\right) ]$ are new
density-dependent parameters. The function $U(s)$\ is evaluated in the
similar way as we have done above. In particular, neglecting the
contribution of plasmons which is again vanishingly small at $s\rightarrow
\infty $ this function becomes
\begin{equation}
U\left( s\right) =\frac{16\alpha ^{2}}{3\pi }\left[ \int_{s-1}^{s}zdz%
\int_{s}^{z+1}L\left( z,u\right)
u^{3}du-\int_{s}^{s+1}zdz\int_{z-1}^{s}L\left( z,u\right) u^{3}du-\frac{\pi s%
}{8\alpha }\right] .
\label{eq:app13}
\end{equation}%
Now only the single-particle excitations contributes to Eq.~(\ref{eq:app13}%
). Again by making the changes of the integration variables, $z\rightarrow
z+s$ and $u\rightarrow u+s$, at $s\rightarrow \infty $ we have found that
the function $U(s)$\ behaves as
\begin{equation}
U\left( s\right) =\frac{1}{3}-\varkappa _{1}\left( r_{s}\right) +\frac{%
\alpha }{6s}+\mathrm{O}\left( s^{-2}\right) .
\label{eq:app14}
\end{equation}%
Finally, substituting Eq.~(\ref{eq:app14}) into Eq.~(\ref{eq:app12}) and
taking the limit $s\rightarrow \infty $ we arrive at
\begin{equation}
\ln h_{2}\left( r_{s}\right) =\frac{1}{3\left[ 1-\varkappa _{1}\left(
r_{s}\right) \right] }\left\{ \ln \left( \frac{2\sqrt{2}}{r_{s}}\right)
+3\varkappa _{1}\left( r_{s}\right) \left[ 1+\ln \varkappa _{2}\left(
r_{s}\right) \right] -2\right\} .
\label{eq:app15}
\end{equation}%
The transition to the limit of non-interacting 2D electron gas is performed
by taking the limit $\varkappa _{1}\rightarrow 0$ in Eq.~(\ref{eq:app15})
which yield
\begin{equation}
\ln h_{2}\left( r_{s}\right) =\frac{1}{3}\left[ \ln \left( \frac{2\sqrt{2}}{%
r_{s}}\right) -2\right] .
\label{eq:app16}
\end{equation}

\end{document}